\documentclass{aa}

\usepackage{graphicx}
\usepackage{txfonts}
\usepackage[breaklinks=True]{hyperref}
\usepackage{color}
\begin{document} 

   \title{The third data release of the Kilo-Degree Survey and
     associated data products}
   
   \author{Jelte T. A. de Jong\inst{1}
          \and Gijs A. Verdoes Kleijn\inst{2}
          \and Thomas Erben \inst{3}
          \and Hendrik Hildebrandt \inst{3}
          \and Konrad Kuijken \inst{1}
          \and Gert Sikkema \inst{2}
          \and Massimo Brescia \inst{4}
          \and Maciej Bilicki \inst{1,5}
          \and Nicola R. Napolitano \inst{4}
          \and Valeria Amaro \inst{6}
          \and Kor G. Begeman \inst{2}
          \and Danny R. Boxhoorn \inst{2}
          \and Hugo Buddelmeijer \inst{1}
          \and Stefano Cavuoti \inst{4,6}
          \and Fedor Getman \inst{4}
          \and Aniello Grado \inst{4}
          \and Ewout Helmich \inst{2}
          \and Zhuoyi Huang \inst{4}
          \and Nancy Irisarri \inst{1}
          \and Francesco La Barbera \inst{4}
          \and Giuseppe Longo \inst{6}
          \and John P. McFarland \inst{2}
          \and Reiko Nakajima \inst{3}
          \and Maurizio Paolillo \inst{6}
          \and Emanuella Puddu \inst{4}
          \and Mario Radovich \inst{7}
          \and Agatino Rifatto \inst{4}
          \and Crescenzo Tortora \inst{2}
          \and Edwin A. Valentijn \inst{2}
          \and Civita Vellucci \inst{6}
          \and Willem-Jan Vriend \inst{2}
          \and Alexandra Amon \inst{8}
          \and Chris Blake \inst{9}
          \and Ami Choi \inst{8,10}
          \and Ian Fenech Conti \inst{11,12}
          \and Stephen D. J. Gwyn \inst{13}
          \and Ricardo Herbonnet \inst{1}
          \and Catherine Heymans \inst{8}
          \and Henk Hoekstra \inst{1}
          \and Dominik Klaes \inst{3}
          \and Julian Merten \inst{14}
          \and Lance Miller \inst{14}
          \and Peter Schneider \inst{3}
          \and Massimo Viola \inst{1}
          }

   \institute{Leiden Observatory, Leiden University, P.O. Box 9513, 2300 RA Leiden, the Netherlands\\
              \email{jdejong@strw.leidenuniv.nl}
         \and
              Kapteyn Astronomical Institute, University of Groningen, P.O. Box 800, 9700 AV Groningen, the Netherlands
         \and
              Argelander-Institut f\"ur Astronomie, Auf dem H\"ugel 71, D-53121 Bonn, Germany
         \and
              INAF - Osservatorio Astronomico di Capodimonte, Via Moiariello 16 -80131 Napoli, Italy
         \and
              National Centre for Nuclear Research, Astrophysics Division, P.O. Box 447, PL-90-950 \L\'od\'z, Poland
         \and
              Department of Physics "E.Pancini", University Federico II, via Cinthia 6, I-80126 Napoli, Italy
         \and
              INAF - Osservatorio Astronomico di Padova, via dell'Osservatorio 5, 35122 Padova, Italy
         \and
              Scottish Universities Physics Alliance, Institute for Astronomy, University of Edinburgh, Royal Observatory, Blackford Hill, Edinburgh, EH9 3HJ, UK
         \and
              Centre for Astrophysics \& Supercomputing, Swinburne University of Technology, P.O. Box 218, Hawthorn, VIC 3122, Australia
         \and
              Center for Cosmology and AstroParticle Physics, The Ohio State University, 191 West Woodruff Avenue, Columbus, OH 43210, United States
         \and
              Institute of Space Sciences and Astronomy (ISSA), University of Malta, Msida, MSD 2080, Malta 
         \and
              Department of Physics, University of Malta, Msida, MSD 2080, Malta
         \and
              Canadian Astronomy Data Centre, Herzberg Astronomy and Astrophysics, 5071 West Saanich Road, Victoria, BC, V9E 2E7, Canada
         \and
              Department of Physics, University of Oxford, Keble Road, Oxford OX1 3RH, UK
             }

   \date{Received ???; accepted ???}

  \abstract
       {The Kilo-Degree Survey (KiDS) is an ongoing optical wide-field imaging
     survey with the OmegaCAM camera at the VLT Survey Telescope. It
     aims to image 1500 square degrees in four filters ($ugri$). 
     The core science
     driver is mapping the large-scale matter distribution in
     the Universe, using weak lensing
     shear and photometric redshift measurements. Further science
     cases include galaxy evolution, Milky Way structure, 
     detection of high-redshift clusters, and finding rare sources
     such as strong lenses and quasars.}
     {Here we
     present the third public data release and several associated data
     products, adding further area, homogenized photometric calibration,
     photometric redshifts and weak lensing shear measurements to the 
     first two releases.}
     {A dedicated pipeline embedded in the \textsc{Astro-WISE}
     information system is used for the
     production of the main release. Modifications with respect to earlier releases are
     described in detail. Photometric redshifts have been derived using both
     Bayesian template fitting, and machine-learning techniques. For
     the weak lensing measurements, optimized procedures based on the
     \textsc{THELI} data reduction and \textit{lens}fit shear measurement
     packages are used.
     }
     {In this third data release an additional 292 new survey tiles ($\approx 300~ {\rm deg}^2$) stacked $ugri$ images
     are made available, accompanied by weight maps, masks, and source
     lists. The multi-band catalogue, including homogenized
     photometry and photometric redshifts, covers the combined DR1, DR2 and DR3 footprint of 440 survey tiles (447
     ${\rm deg}^2$). Limiting magnitudes are typically 24.3, 25.1, 24.9,
     23.8 (5$\sigma$ in a 2\arcsec aperture) in $ugri$, respectively,
     and the typical $r$-band PSF size is less than
     0.7\arcsec. The photometric homogenization scheme ensures accurate colors and an absolute calibration stable to $\approx 2$\% for $gri$ and $\approx 3$\% in $u$.
     Separately released for the combined area of all KiDS releases to
     date are a weak lensing shear catalogue and 
     photometric redshifts based on two different machine-learning 
     techniques.}
     {}


   \maketitle

  \titlerunning{KiDS data release 3}
  \authorrunning{J. T. A. de Jong et al.}

\section{The Kilo-Degree Survey}
\label{sec:intro}

\begin{figure*}[ht]
   \centering
   \includegraphics[width=\textwidth]{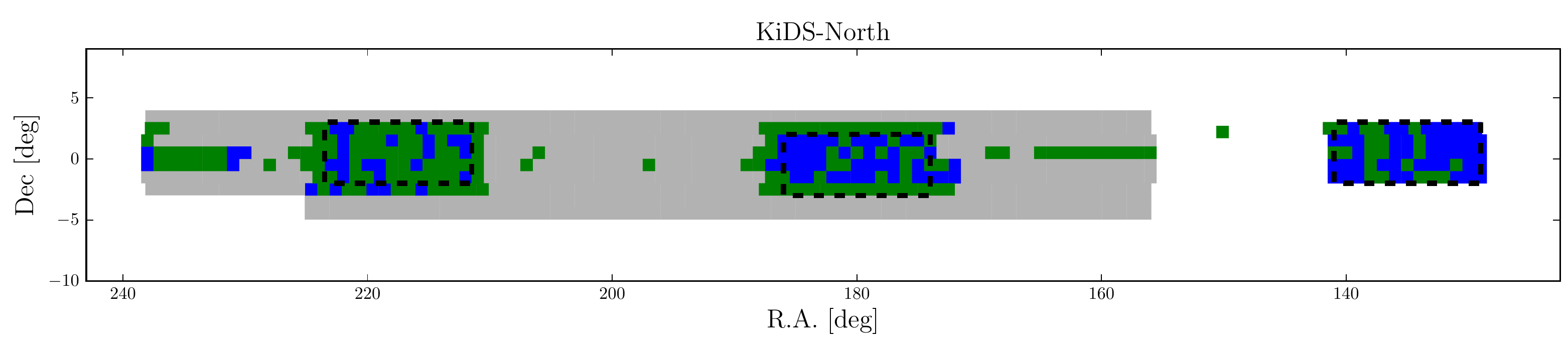}
   \includegraphics[width=\textwidth]{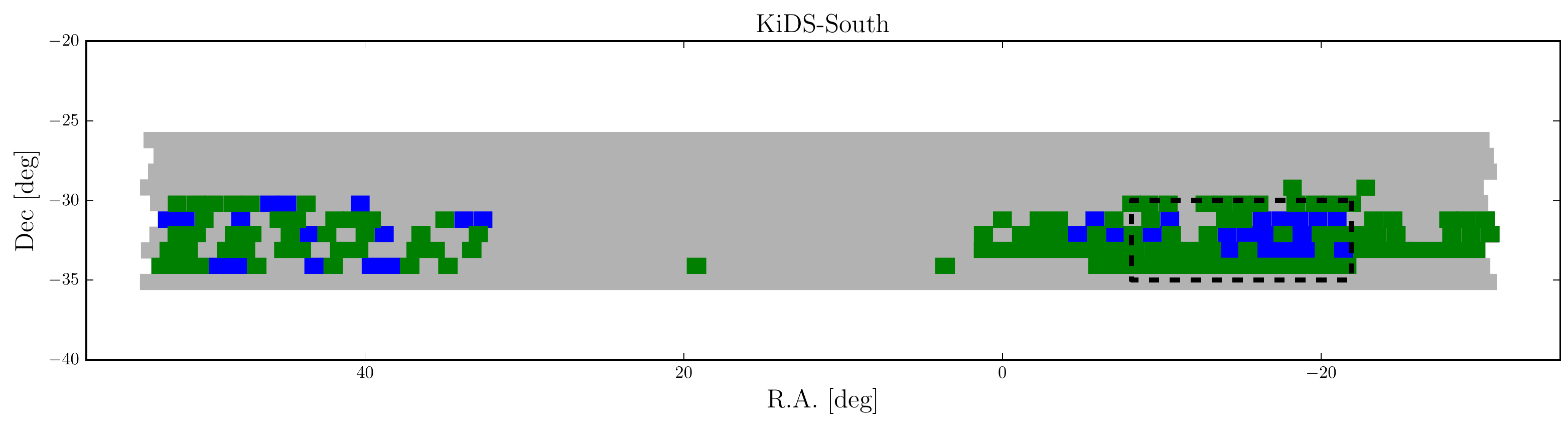}
   \caption{Sky distribution of survey tiles released in KiDS-ESO-DR3
     (green) and in the previous releases KiDS-ESO-DR1 and -DR2
     (blue). The multi-band source catalogue covers the combined area 
     (blue + green) and the full KiDS area is shown in grey. 
     {\it Top:} KiDS-North. {\it Bottom:} KiDS-South. Black dashed 
     lines delineate the locations of the GAMA fields; the single
     pointing at RA=150\degr is centered at the COSMOS/CFHTLS D2 field.}
              \label{Fig:SkyDistribution}
    \end{figure*}

With the advent of specialized wide-field telescopes and cameras,
large multi-wavelength astronomical imaging surveys have become
important tools for astrophysics. In addition to the specific scientific goals that they are designed
and built for, their data have huge legacy value and facilitate a
large variety of scientific analyses. For these reasons, since their
commissioning in 2011 the VLT
Survey Telescope \cite[VST, ][]{capaccioli/etal:2012} and its sole instrument, the 268 Megapixel
OmegaCAM camera \citep{kuijken:2011} have been mostly dedicated to three large public
surveys\footnote{\url{http://www.eso.org/sci/observing/PublicSurveys.html}},
of which the Kilo-Degree Survey \citep[KiDS,][]{dejong/etal:2013} is the largest
in terms of observing time. It aims to observe 1500 ${\rm deg}^2$ of
extragalactic sky, spread over two
survey fields, in four broad-band filters ($ugri$)  (see Fig.~\ref{Fig:SkyDistribution} and Table
\ref{Tab:fields}). Together with
its sister survey, the VISTA Kilo-Degree Infrared Galaxy
Survey \citep[VIKING, ][]{edge/etal:2013}, this will result in a 9-band
optical-infrared data set with excellent depth and image
quality. 

KiDS was designed primarily to serve as a weak gravitational lensing
tomography survey, mapping the large-scale matter distribution in the
Universe. Key requirements for this application are unbiased and
accurate weak lensing shear measurements and photometric redshifts,
which put high demands on both image quality and depth, as well as the
calibration. The VST-OmegaCAM system is ideal for such a survey, as
it was specifically designed to provide superb and uniform image
quality over a large, 1\degr$\times$1\degr, field of view
(FOV). Combining a science array of 32 thinned, low-noise 2k$\times$4k
E2V devices, a constant
pixel scale of 0.21\arcsec, and real-time wave-front sensing and
active optics, the system provides a PSF equal to the atmospheric
seeing over the full FOV down to 0.6\arcsec. To achieve optimal shear
measurements, the survey makes use of the flexibility of service mode
scheduling. Observations in the $r$ band are taken under the
best dark-time conditions, with $g$ and $u$ in increasingly worse
seeing. The $i$ band is the only filter observed in bright time. The
observing condition constraints for execution of KiDS OBs are listed
in Table \ref{Tab:ObservingConstraints}. Apart from the primary
science goal, the KiDS data are exploited for a large range of
secondary science cases, including quasar, strong gravitational lens
and galaxy cluster searches, galaxy evolution, studying the matter
distribution in galaxies, groups and clusters, and even Milky Way
structure \cite[e.g.][]{dejong/etal:2013}.

\begin{table*}
\caption{KiDS observing strategy: observing condition constraints and exposure times.}
\label{Tab:ObservingConstraints}
\centering
\begin{tabular}{l c c c c c c c}
\hline\hline
Filter & Max. lunar & Min. moon & Max. seeing & Max. airmass & Sky transp. & Dithers & Total Exp.\\
~ & illumination & distance (deg) & (arcsec) & ~ & ~ & ~ & time (s) \\
\hline
$u$ & 0.4 & 90 & 1.1 & 1.2 & CLEAR & 4 & 1000 \\
$g$ & 0.4 & 80 & 0.9 & 1.6 & CLEAR & 5 & 900 \\
$r$ & 0.4 & 60 & 0.8 & 1.3 & CLEAR & 5 & 1800 \\
$i$ & 1.0 & 60 & 1.1 & 2.0 & CLEAR & 5 & 1200 \\
\hline
\end{tabular}
\end{table*}

The first two data releases from KiDS \citep{dejong/etal:2015} became public in
2013 and 2015, containing reduced image data, source lists and a
multi-band catalog for a total of 148 survey tiles ($\simeq$160
${\rm deg}^2$). Based on this data set, the first scientific results focused
on weak lensing studies of galaxies and galaxy groups in the Galaxy
And Mass Assembly \citep[GAMA, ][]{driver/etal:2011} fields
\citep{kuijken/etal:2015,sifon/etal:2015,viola/etal:2015,brouwer/etal:2016,vanuitert/etal:2016}.
Furthermore, these data releases were the basis for  
the first results from a $z\sim 6$ quasar search
\citep{venemans/etal:2015}, a catalogue of photometric redshifts from machine-learning
\citep{cavuoti/etal:2015,cavuoti/etal:2017}, preliminary results on super-compact massive
galaxies \citep{tortora/etal:2016}, and a catalog of galaxy clusters \citep{radovich/etal:2017}.

In this publication we present the third KiDS data release
(KiDS-ESO-DR3). Extending the total released data set to 440 survey
tiles (approximately 450 ${\rm deg}^2$.), this release also includes photometric redshifts and a global
photometric calibration. In addition to the core ESO release,
several associated data products are described that have been
released. These include
photo-z probability distribution functions, machine-learning
photo-z's, weak lensing shear catalogs and lensing-optimized
image data. The first applications of this new data
set have appeared already and include the first cosmological results from KiDS 
\citep{joudaki/etal:2016,brouwer/etal:2017,hildebrandt/etal:2017}.

The outline of this paper is as follows. Sect. \ref{sec:dr3} is a
discussion of the contents of KiDS-ESO-DR3 and the differences in terms
of processing and data products with respect to earlier releases. Sect. 
\ref{sec:weaklensing} presents the weak lensing data products and Sect. 
\ref{sec:photz} reviews the different sets of photometric redshifts that
are made available. Data access routes are summarized in Sect. 
\ref{sec:access} and a summary and outlook towards future data releases 
is provided in Sect. \ref{sec:summary}.

\begin{table}[ht]
\caption{KiDS Fields (see also Fig.~\ref{Fig:SkyDistribution}).}
\label{Tab:fields}
\centering
\begin{tabular}{llll}
\hline\hline
Field & RA range & Dec range & Area \\
\hline\noalign{\smallskip}
KiDS-S    & 22$^{\rm h}$00$^{\rm m}$ -- 3$^{\rm h}$30$^{\rm m}$ & $-$36\degr -- $-$26\degr & 720 deg$^2$ \\
\noalign{\smallskip}\hline\noalign{\smallskip}
KiDS-N    & 10$^{\rm h}$24$^{\rm m}$ -- 15$^{\rm h}$00$^{\rm m}$ & $-$5\degr -- $+$4\degr & 712 deg$^2$ \\
~         & 15$^{\rm h}$00$^{\rm m}$ -- 15$^{\rm h}$52$^{\rm m}$ & $-$3\degr -- $+$4\degr & ~ \\
\noalign{\smallskip}\hline\noalign{\smallskip}
KiDS-N-W2 & 8$^{\rm h}$30$^{\rm m}$ -- 9$^{\rm h}$30$^{\rm m}$ & $-$2\degr -- $+$3\degr & 68 deg$^2$ \\
\noalign{\smallskip}\hline\noalign{\smallskip}
KiDS-N-D2 & 9$^{\rm h}$58$^{\rm m}$ -- 10$^{\rm h}$02$^{\rm m}$ & $+$1.7\degr -- $+$2.7\degr & 1 deg$^2$ \\
\hline
\end{tabular}
\end{table}

\section{The third data release}
\label{sec:dr3}

\subsection{Content and data quality}
\label{sec:content}

KiDS-ESO-DR3 (DR3) constitutes the third data release of KiDS and can be considered an incremental release that adds area, an improved photometric calibration, and photometric redshifts to the two previous public data releases.

In its approximately yearly data releases, KiDS provides data products
for survey tiles that have been successfully
observed in all four filters ($u$, $g$, $r$, $i$). Adding 292 new
survey tiles to the 50 (DR1) and 98 (DR2) already released tiles, DR3
brings the total released area to 440 tiles. DR3 includes complete
coverage of the Northern GAMA fields, which were targeted first, in
order to maximize synergy with this spectroscopic survey early on. The
distribution of released tiles on the sky is shown in
Fig.~\ref{Fig:SkyDistribution} and a complete list, including data
quality information, can be found on the KiDS DR3
website\footnote{\url{http://kids.strw.leidenuniv.nl/DR3}}. For these
292 tiles the following data products are provided for each filter (as
they were for DR1 and DR2):
\begin{itemize}
  \item astrometrically and photometrically calibrated, stacked images (``coadds'')
  \item weight maps
  \item flag maps (``masks''), that flag saturated pixels, reflection halos, read-out spikes, etc.
  \item single-band source lists
\end{itemize}

Slight gain variations exist across the FOV, but an average effective gain for each coadd is provided in the tile table on the KiDS DR3 website.
The final calibrated, coadded images have a uniform pixel scale of
0.2\arcsec, and the pixel values are in units of flux relative to the flux
corresponding to magnitude = 0. The magnitude $m$ corresponding to a
pixel value $f$ is therefore given by:
\begin{equation}
\label{Eq:fluxscale}
m = -2.5 \log_{10} f.
\end{equation}

The single-band source lists are identical in format and content to those 
released in DR2 and contain an extensive set of SExtactor \citep{bertin/arnouts:1996} based magnitude and
geometric measurements, to increase their versatility for the end
user. For example, the large number (27) of aperture magnitudes allows
users to use interpolation methods (e.g. ``curve of growth'') to
derive their own aperture corrections or total magnitudes. Also
provided are a star-galaxy separation parameter and information on the
mask regions that might affect individual source measurements. Table
\ref{Tab:singlebandcolumns} lists the columns that are present in the
single-band source lists provided in KiDS-ESO-DR3.

An aperture-matched multi-band catalog is also part of DR3. This
catalog covers not only the 292 newly released tiles, but also the 148
tiles released in DR1 and DR2, for a total of 48\,736\,591 sources in 440 tiles and an area of
approximately 447 square degrees. Source detection, positions and shape parameters are all based on the
$r$-band images, since these typically have the best image quality,
thus providing the most reliable measurements. 
The star-galaxy separation provided is the same as that in the
$r$-band single-band source list, and is based on separating point-like from extended sources, which in some tiles yields sub-optimal results due to PSF variations. 
In future releases we plan to include more information, for example from colors or PSF models, to improve this classification.
Magnitudes are measured
in all filters using forced photometry. Seeing differences are
mitigated in two ways: 1) via aperture-corrected magnitudes, and 2)
via Gaussian Aperture and PSF (GAaP) photometry. The latter is new in DR3 and described in
more detail in Sect. \ref{sec:gaap}. In this release we also introduce a new photometric calibration scheme, based on the GAaP measurements, that homogenizes the photometry in the catalogue over the survey area,
using a combination of stellar locus information and overlaps between
tiles. The procedures used and the quality of the results are reviewed
in Sect. \ref{sec:phothom}. Also new in the multi-band catalog with
respect to DR2 are photometric redshifts, see
Sect. \ref{sec:photz-bpz}. All columns available in this catalog are
listed in Table \ref{Tab:MultiBandColumns}.

   \begin{figure*}[ht]
   \centering
   \includegraphics[width=\textwidth]{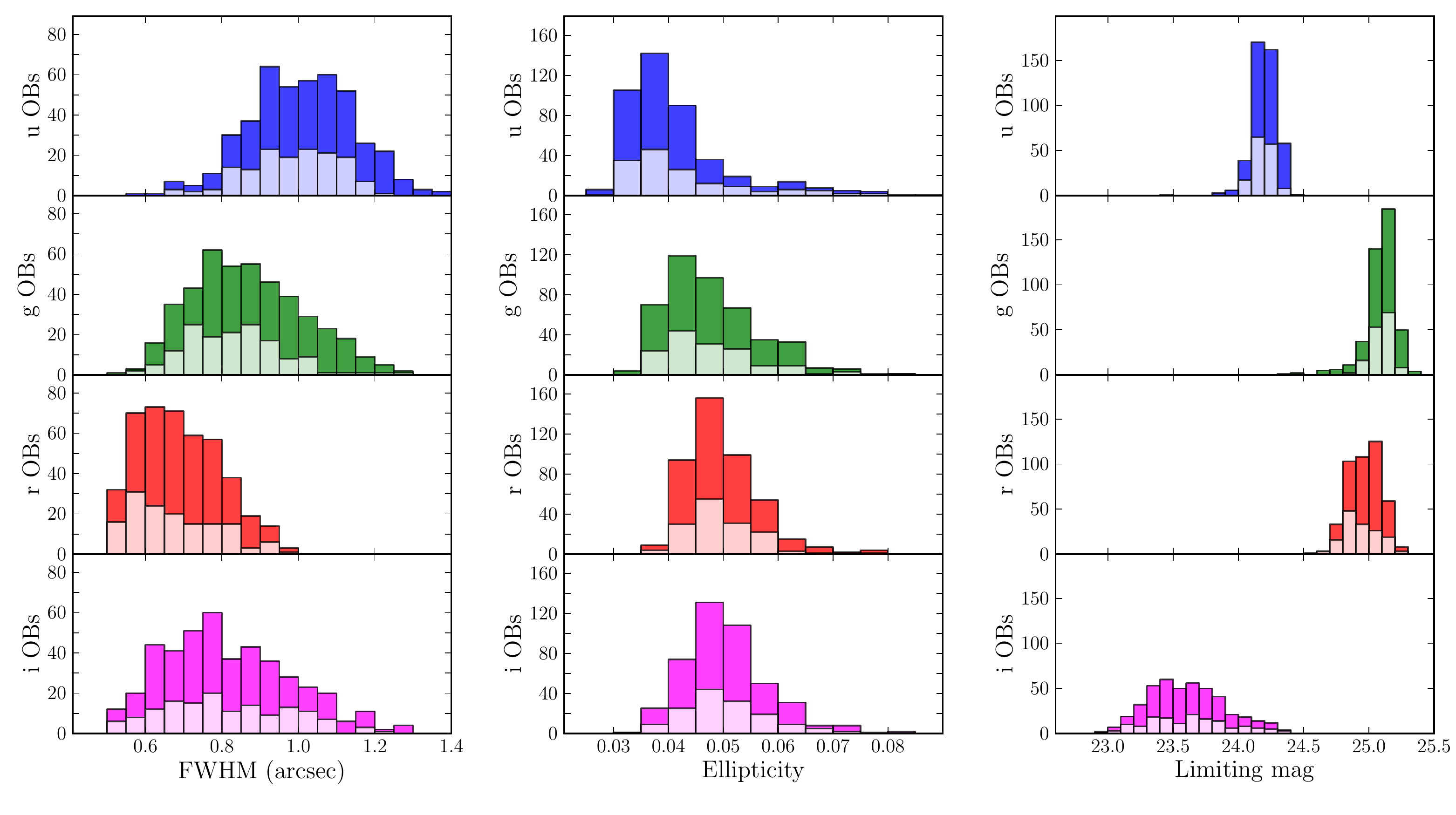}
   \caption{Data quality for KiDS-ESO-DR1 -DR2 and -DR3. {\it Left:}
     average PSF size (FWHM) distributions; {\it center:} average PSF
     ellipticity distributions; {\it right:} limiting magnitude
     distributions (5$\sigma$ AB in 2\arcsec\ aperture). The
     distributions are per filter: from top to bottom $u$, $g$, $r$,
     and $i$, respectively. The full histograms correspond to the 440 
     tiles included in the DR3 multi-band catalog, while the lighter 
     portions of the histograms correspond to fraction (148 tiles) 
     previously released in KiDS-ESO-DR1 and -DR2.} 
              \label{Fig:DataQuality}
    \end{figure*}

\begin{table}
\centering
\caption{Data quality of all released tiles}
\label{Tab:DataQuality}
\begin{tabular}{l c c c c c c}
\hline\hline
Filter & \multicolumn{2}{c}{PSF FWHM} & \multicolumn{2}{c}{PSF Ellipticity} & \multicolumn{2}{c}{Limiting mag.}\\
~ & \multicolumn{2}{c}{(\arcsec)} & \multicolumn{2}{c}{(1-$b$/$a$)} &
\multicolumn{2}{c}{(5$\sigma$ in 2\arcsec ap.)} \\
~ & Mean & Scatter & Mean & Scatter & Mean & Scatter \\
\hline
$u$ & 1.00 & 0.13 & 0.041 & 0.010 & 24.20 & 0.09  \\
$g$ & 0.86 & 0.14 & 0.047 & 0.008 & 25.09 & 0.11 \\
$r$ & 0.68 & 0.11 & 0.049 & 0.006 & 24.96 & 0.12 \\
$i$ & 0.81 & 0.16 & 0.050 & 0.008 & 23.62 & 0.30 \\
\hline\\
\end{tabular}
\end{table}

The intrinsic data quality of the KiDS data is illustrated in
Fig. \ref{Fig:DataQuality} and summarized in Table
\ref{Tab:DataQuality}. Average image quality, quantified by the size
(FWHM in arcsec) of the point-spread-function (PSF), is driven by the
observing constraints supplied to the scheduling system (Table
\ref{Tab:ObservingConstraints}). The aim here is to use the best dark
conditions for $r$-band, which is used for the weak lensing shear
measurements, with increasingly worse seeing during dark-time
allocated to $g$ and $u$, respectively. The only filter observed in
grey and bright time, $i$-band makes use of a large range of seeing
conditions. To improve (relative) data rates in the dark-time filters,
the seeing constraints have been relaxed slightly. So far this
has not resulted in a significant detrimental effect on the
overall image quality of the new DR3 data, when compared to the DR1
and DR2 data. Average ellipticities of stars over the FOV (middle column of
Fig. \ref{Fig:DataQuality}), here defined as $1-b/a$ and measured by
SExtractor \citep{bertin/arnouts:1996}, are always significantly
smaller than 0.1. The depth of the survey is quantified by a
signal-to-noise of 5$\sigma$ for point sources in 2\arcsec
apertures. Despite slightly poorer average seeing the $g$-band data are
marginally deeper than the $r$-band data. The large range of limiting
magnitudes in $i$-band reflects the variety in both seeing and sky
illumination conditions. The overall data quality of the DR3 release
is very similar to the data quality of DR1 and DR2, as described in
\cite{dejong/etal:2015} and \cite{kuijken/etal:2015}.

\begin{table*}
\caption{VST baffle configurations}
\label{Tab:BafflingConfigurations}
\centering
\begin{tabular}{clcc}
\hline\hline
Config. & Description & Start date & End date \\
\hline
1 & Original set-up & May 2011 & 7 Jan 2014 \\
2 & Baffle extensions 1 (M2, chimney) & 8 Jan 2014 & 6 Apr 2014 \\
3 & Baffle extensions 2 (M2, chimney, M1 plug) & 7 Apr 2014 & 29 Apr 2015 \\
4 & Baffle extensions 3 (chimney ridges, plug) & 30 Apr 2015 & present \\
\hline
\end{tabular}
\end{table*}

The most striking and serious issues with the KiDS data are caused by
stray light that scatters into the light path and onto the focal plane
\citep[see][for some examples]{dejong/etal:2015}. 
Over the course of 2014 and 2015, the VST baffles were significantly
redesigned and improved (see Table \ref{Tab:BafflingConfigurations}). 
As a result, many of the stray light issues
that affect the VST data are now much reduced or eliminated. Although a fraction of
the DR3 observations were obtained with
improved telescope baffles, the majority of the $i$-band data, which
is most commonly affected, was obtained with the original
configuration. Severely affected images are flagged in the tile and
catalog tables on the KiDS DR3 website, and sources in affected tiles
are flagged in the multi-band catalog included in the ESO release.

\subsection{Differences with DR1 and DR2}
\label{sec:differences}

Data processing for KiDS-ESO-DR3 is
based on a KiDS-optimized version of the {\sc Astro-WISE} optical pipeline described in
\cite{mcfarland/etal:2013}, combined with dedicated masking and source extraction procedures. The pipeline and procedures used are largely identical to those used for DR2, and for a detailed discussion we refer to \cite{dejong/etal:2015}. In the following sections only the differences and additional procedures are described in detail.

\subsubsection{Pixel processing}

\noindent {\bf Cross-talk correction.}
Data processed for DR3 were observed between the 9th of August 2011
and the 4th of October 2015.  Since the electronic cross-talk between
CCDs \#95 and \#96 is stable for certain periods, these stable
intervals had to be determined for the period following the last
observations processed for the earlier releases.  The complete set of
stable periods and the corrections applied are listed in Table
\ref{Tab:CrosstalkCoefficients}.

\begin{table}
\caption{Applied cross-talk coefficients.}
\label{Tab:CrosstalkCoefficients}
\centering
\footnotesize
\resizebox{\columnwidth}{!}{
\begin{tabular}{l | c c | c c}
\hline\hline
Period & \multicolumn{2}{c|}{CCD \#95 to CCD \#96\tablefootmark{a}} & \multicolumn{2}{c}{CCD \#96 to CCD \#95\tablefootmark{a}} \\
~ & $a$ & $b$ ($\times10^{-3}$) & $a$ & $b$ ($\times10^{-3}$)\\
\hline
2011-08-01 - 2011-09-17 & $-$210.1 & $-$2.504 & 59.44 & 0.274 \\
2011-09-17 - 2011-12-23 & $-$413.1 & $-$6.879 & 234.8 & 2.728 \\
2011-12-23 - 2012-01-05 & $-$268.0 & $-$5.153 & 154.3 & 1.225 \\
2012-01-05 - 2012-07-14 & $-$499.9 & $-$7.836 & 248.9 & 3.110 \\
2012-07-14 - 2012-11-24 & $-$450.9 & $-$6.932 & 220.7 & 2.534 \\
2012-11-24 - 2013-01-09 & $-$493.1 & $-$7.231 & 230.3 & 2.722 \\
2013-01-09 - 2013-01-31 & $-$554.2 & $-$7.520 & 211.9 & 2.609 \\
2013-01-31 - 2013-05-10 & $-$483.7 & $-$7.074 & 224.7 & 2.628 \\
2013-05-10 - 2013-06-24 & $-$479.1 & $-$6.979 & 221.1 & 2.638 \\
2013-06-24 - 2013-07-14 & $-$570.0 & $-$7.711 & 228.9 & 2.839 \\
2013-07-14 - 2014-01-01 & $-$535.6 & $-$7.498 & 218.9 & 2.701 \\
2014-01-01 - 2014-03-08 & $-$502.2 & $-$7.119 & 211.6 & 2.429 \\
2014-03-08 - 2014-04-12 & $-$565.8 & $-$7.518 & 215.1 & 2.578 \\
2014-04-12 - 2014-08-12 & $-$485.1 & $-$6.887 & 201.6 & 2.237 \\
2014-08-12 - 2014-01-09 & $-$557.9 & $-$7.508 & 204.2 & 2.304 \\
2014-01-09 - 2015-05-01 & $-$542.5 & $-$7.581 & 219.9 & 2.535 \\
2015-05-01 - 2015-07-25 & $-$439.3 & $-$6.954 & 221.5 & 2.395 \\
2015-07-25 - 2015-08-25 & $-$505.6 & $-$7.535 & 229.7 & 2.605 \\
2015-08-25 - 2015-11-10 & $-$475.2 & $-$7.399 & 218.0 & 2.445 \\
\hline
\end{tabular}
}
\tablefoot{
\tablefoottext{a}{
Correction factors $a$ and $b$ are applied to each pixel in target CCD based on the pixel values in the source CCD:
\begin{equation}
I'_i = 
\begin{cases}
I_i + a, &\text{if $I_j = I_{\rm{sat.}}$;}\\
I_i + b  I_j, &\text{if $I_j < I_{\rm{sat.}}$,}\\
\end{cases}
\end{equation}
where $I_i$ and $I_j$ are the pixel values in CCDs $i$ and $j$, $I'_i$ is the corrected pixel value in CCD $i$ due to cross-talk from CCD $j$, and $I_{\rm{sat.}}$ is the saturation pixel value.
}
}
\end{table}

\noindent {\bf Flatfields and illumination correction.}
The stray light issues in VST that were addressed with changes to the
telescope baffles in 2014 and 2015 do not only affect the science
data, but also flatfields. Such additional light present in the flat
field results in non-uniform illumination and must be corrected by an
"illumination correction" step. Because the illumination of the focal
plane changed for each baffle configuration (Table
\ref{Tab:BafflingConfigurations}), new flat fields and associated
illumination corrections are required for each configuration. Thus,
whereas for DR1 and DR2 a single set of masterflats was used for each
filter, new masterflats were created for each of the baffle
configurations. The stability of the intrinsic pixel
sensitivities\footnote{constant to ~0.2\% or better for g, r and i
  \citep{verdoes/etal:2013,dejong/etal:2015}} still allows a single set to be used for
each configuration. Our method to derive the illumination correction
makes use of specific calibration observations where the same
standard stars are observed with all 32 CCDs \cite[see][]{verdoes/etal:2013}. 
Because such data are not available for all configurations, 
the differences between the original flatfields and those for each
baffle configuration were used to derive new illumination corrections from
the original correction.

\subsubsection{Masking}

Bright stars and related features, such
as read-out spikes, diffraction spikes and reflection halos, are
masked in the newly released tiles by the Pulecenella
software \cite[see][]{dejong/etal:2015}. In prior releases these automated
masks were complemented with manual masks that covered a range of
other features, including stray light, remaining satellite tracks and
reflections/shadows of the covers over the detector heads. 
Because this way of masking is inherently subjective and not
reproducible, and given the considerable effort required, for DR3 we
do not provide manual masks. 
The fraction of masked area in the 292 new tiles is 1.5\%, 5.8\%, 14.6\% and 10.1\% in $u$, $g$, $r$ and $i$, respectively. In the 148 tiles released earlier the fraction of automatically masked area was very similar, but the manual masks added a further 2\%, 7\%, 10\% and 14\% to this, effectively doubling the total masked area. Using the mask flag values the manual masks in the DR1 and DR2 tiles can be ignored in order to obtain consistent results over the full survey area.

To replace the manually created masks, an automated procedure is under 
development that aims to identify the same types of areas. This
procedure, dubbed MASCS+, will be described in, and the resulting masks
released jointly with, a forthcoming paper (Napolitano et al. in
prep).

\subsubsection{Photometry and redshifts}

The main enhancements of KiDS-ESO-DR3 over DR1 and DR2 are the
improved photometric calibration and the inclusion of GAaP \citep[see][]{kuijken/etal:2015}
measurements and photometric redshifts.  Where in the earlier releases
the released tiles formed a very patchy on-sky distribution, the
combined set of 440 survey tiles now available mostly comprises a
small number of large contiguous areas, allowing a refinement of the
photometric calibration that exploits both the overlap between
observations within a filter as well as the stellar colours across
filters. This photometric homogenization scheme is the subject of
Sect. \ref{sec:phothom}.

GAaP is a two-step procedure that homogenizes the PSF over the full
FOV of a survey tile, and measures a seeing-independent magnitude in a
Gaussian-weighted aperture. This type of measurement yields accurate
galaxy colours and approaches PSF-fitting photometry for point
sources. See Sect. \ref{sec:gaap} for more details. The colours are
used as input for the photometric redshifts included in the DR3
catalog (Sect. \ref{sec:photz}), as well as for the machine-learning
based photometric redshifts discussed in Sect. \ref{sec:photz}.

\subsection{Additional catalog columns}

Compared to the multi-band catalog released with KiDS-ESO-DR2, the
current multi-band catalog contains a number of additional columns:

\noindent{\bf Extinction.} For every source the foreground extinction
is provided. The colour excess $E(B-V)$ at the source position is
transformed to the absorption $A$ in each filter, based on the maps
and coefficients by \cite{schlegel/etal:1998}. These extinctions can be directly
applied to the magnitudes in the catalog, which are {\it not
  corrected} for extinction.

\noindent{\bf Tile quality flag.} In the absence of manual masking of
severe image defects, sources in survey tiles with one or more poor
quality coadded images (defined as such during visual inspection of
all images) are flagged\footnote{This information is also available in
the tile table on the DR3 website (\url{http://kids.strw.leidenuniv.nl/DR3})}. The bitmap value indicates which
filter is affected: 1 for $u$, 2 for $g$, 4 for $r$, and 8 for $i$.

\noindent{\bf GAaP magnitudes and colours.} For each filter the GAaP
magnitude for each source is provided, together with the error
estimate. The aperture is defined from the $r$-band image (see
Sect. \ref{sec:gaap}). Also included are six colours, based on the GAaP
magnitudes.

\noindent{\bf Photometric homogenization.} The photometric
homogenization procedure (Sect. \ref{sec:phothom}) results in
zeropoint offsets for each filter in each survey tile. These offsets
are included in separate columns and can be applied to the magnitude
columns, which are {\it not homogenized}. Since the homogenization is
based on the GAaP magnitudes, care should be taken when applying these
offsets to other magnitude measurements (see Sect. \ref{sec:phothom}).

\noindent{\bf Photometric redshifts.} Results from the application of
{\sc BPZ} \citep{benitez:2000} to the homogenized and extinction-corrected GAaP magnitudes
are provided in three columns: i) the best-fitting photometric
redshift, ii) the ODDS (Bayesian odds) and iii) the best-fitting spectral
template (see Sect. \ref{sec:photz} for details). 

\noindent{\bf Astro-WISE identifiers.} To enable straightforward
cross-matching, and tracing of data lineage with {\sc Astro-WISE}, the
identifiers of the SourceCollections, SourceLists and individual
sources therein are propagated.

\subsection{Gaussian Aperture and PSF Photometry}
\label{sec:gaap}

For some applications, in particular photometric redshifts, reliable
colours of each source are needed. For this purpose we provide
`Gaussian Aperture and PSF' photometry.  These fluxes are
defined as the Gaussian-aperture weighted flux of the intrinsic
(i.e. not convolved with the seeing PSF) source $f(x,y)$, with the
aperture defined by its major and minor axis lengths $A$ and $B$, and
its orientation $\theta$:
\begin{equation}
F_{\rm GAaP}=\int\!\!\int {\rm d}x\,{\rm d}y\, f(x,y) {\rm e}^{-[(x'/A)^2+(y'/B)^2]/2}
\end{equation} 
where $x'$ and $y'$ are coordinates rotated by an angle $\theta$ with
respect to the $x$ and $y$ axes.

When the PSF is Gaussian, with rms radius $p$, $F_{\rm GAaP}$ can be
related directly to the Gaussian-aperture flux with axis lengths
$\sqrt{A^2-p^2}$ and $\sqrt{B^2-p^2}$, measured on the
seeing-convolved image. It thus provides a straightforward way to
compensate for seeing differences between different images, and obtain
aperture fluxes in different bands that are directly comparable (in
the sense that each part of the source gets the same weight in the
different bands).

Achieving a Gaussian PSF is done via construction of a convolution
kernel that varies across the image, modelled in terms of shapelets
\citep{refregier:2003}. The size of the Gaussian PSF is set so as to
preserve the seeing as much as possible without deconvolving (which
would amplify the noise). The resulting correlation of the pixel noise
is propagated into the error estimate on $F_{\rm GAaP}$. Full details
of this procedure are provided in \citet[Appendix
  A]{kuijken/etal:2015}.

It is important to note that the GAaP fluxes are primarily intended
for colour measurements: because the Gaussian aperture function tapers
off they are \emph{not} total fluxes (except for point sources). As a
rule of thumb, for optimal SNR it is best to choose a GAaP aperture
that is aligned with the source, and with major and minor axis length
somewhat larger than $\sqrt{A_{\rm obs}^2+p^2}$ and $\sqrt{B_{\rm
    obs}^2+p^2}$. In the KiDS-ESO-DR3 catalogues we set $A$ and $B$ by
adding 0.7 arcsec in quadrature to the measured $r$-band rms major and
minor axis radii, with a maximum of 2 arcsec.

\subsection{Photometric homogenization}
\label{sec:phothom}

KiDS-ESO-DR3 contains zeropoint corrections for all 440 tiles (1760
coadds) to correct for photometric offsets, for instance due to
changes in atmospheric transparency, non-availability of standard star
observations during the night (in which case a default is used) or
other deviations. The corrections are based on a combination of two
methods. The first method is based on the overlaps of adjacent coadds
for a particular passband, which we refer to as Overlap Photometry
(OP). Out of the 440 tiles, 421 are part of a connected group of 10 or
more survey tiles. The second method is a form of stellar locus
regression (SLR) and is based on the $u,g,r,i$ colour information of
each tile.

The best OP results, determined from comparisons to SDSS DR9
\citep{ahn/etal:2012} stellar photometry, are obtained for the
$r$-band. SLR works well for $g$, $r$ and $i$, but delivers poor
results in $u$-band. Also, SLR provides colour calibration, but no
absolute calibration. For these reasons a combination of OP and SLR is
used for the overall photometric homogenization. OP is used for a
homogeneous, absolute $r$-band calibration, to which $g$- and $i$-band
are tied using SLR. For the $u$-band we solely use an independent OP
solution.

\subsubsection{Overlap photometry}

\begin{figure}
   \centering
   \includegraphics[width=\columnwidth]{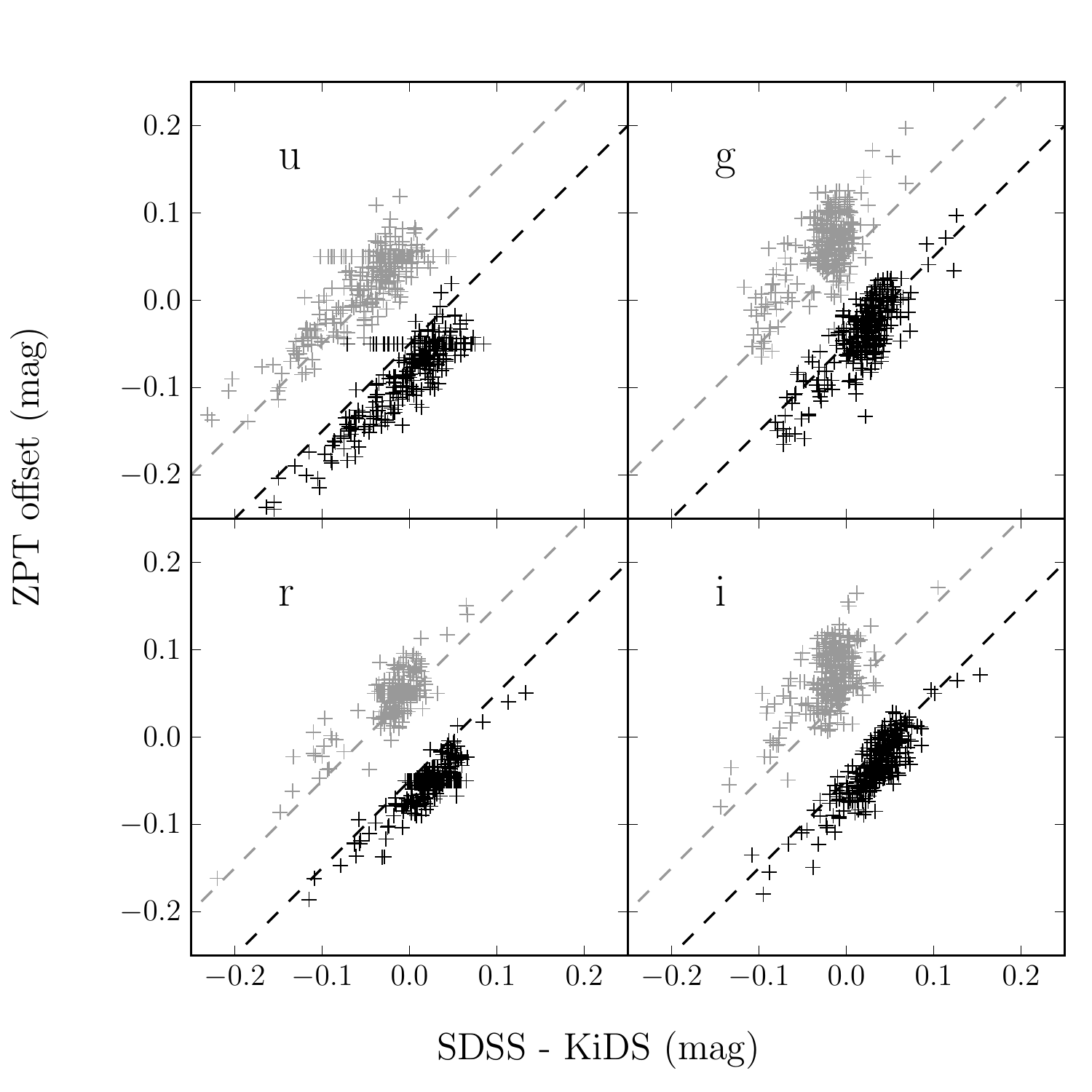}
   \caption{Photometric homogenization zeropoint offsets vs. offsets
     to SDSS. The offsets calculated by the photometric homogenization
   procedure are plotted as function of the per-tile offsets of
   stellar photometry between tiles in KiDS-North and SDSS DR9. Black
   and grey symbols show the SDSS offsets using GAaP and 10\arcsec~
   aperture-corrected photometry, respectively, with the dashed lines
   indicating equality. The GAaP and aperture-corrected data are
   shifted down and up by 0.05 magnitude, respectively, to improve the
   clarity of the figure. Each subpanel corresponds to a different
   passband, denoted by the labels.}
   \label{Fig:HomVsSdss}
\end{figure}

This method involves homogenizing the calibration within a single
passband using overlapping regions of adjacent coadds. Neighbouring
coadds have an overlap of 5\% in RA and 10\% in DEC. The OP
calculations used in this release are based on the GAaP photometry
(see Sect. \ref{sec:gaap}) of point-like objects\footnote{selected
  here with CLASS\_STAR > 0.8}.  OP can only be used if a tile has
sufficient overlap with at least one other tile. This is the case for
431 out of the 440 tiles, leaving 9 isolated tiles (singletons). These
431 tiles are divided over 10 connected groups, which contain 2, 3, 
5,10, 20, 50, 65, 89, 92 and 95 members, respectively.

Photometric anchors, observations with reliable photometric
calibration, are defined and all other tiles are tied to these. The
anchors are selected based on a set of four criteria, that were
established based on a comparison of KiDS GAaP magnitudes with SDSS
DR9 psfMag measurements. This comparison is limited to the KiDS-North
field, since KiDS-South does not overlap with the SDSS footprint.
The following four criteria, that depend only
on the KiDS data, were found to minimize the magnitude offsets between
KiDS and SDSS:
\begin{enumerate}
\item initial calibration based on nightly standard star observations
\item <0.2\arcsec difference in PSF FWHM with the nightly standard star observation
\item observed after April 2012 (following the replacement of a faulty video board\footnote{see the OmegaCAM news page for more details: https://www.eso.org/sci/facilities/paranal/instruments/omegacam/news.html})
\item <0.02 mag atmospheric extinction difference between the exposures
\end{enumerate}
Based on these criteria, in the $r$-band 45\% of all tiles are anchors
and in the $u$-band 18\% are.  A fitting algorithm is applied to each
group and filter independently to minimize the zeropoint differences
in the overlaps, with the constraint that the anchor zeropoints should
not be changed. Tiles that are singletons receive no absolute calibration 
correction in either $u$- or $r$-band.

\subsubsection{Stellar locus regression}

The majority of stars form a tight sequence in colour-colour space,
the so-called `stellar locus'. Inaccuracies in the photometric
calibration of different passbands will shift the location of this
stellar locus. Changing the photometric zeropoints such that the
stellar locus moves to its intrinsic location should correct these
calibration issues. Of course, this procedure only corrects the
zeropoints relative to each other, thus providing colour calibration
but no absolute calibration. In KiDS-ESO-DR3 SLR is used to calibrate
the $g-r$, and $r-i$ colours which, together with the absolute
calibration of the $r$-band using OP, yields the calibration for $g$
and $i$. 

The intrinsic location of the stellar locus is defined by the
`principal colours' derived by \cite{ivezic/etal:2004}. In these
principal colours, linear combinations of $u$, $g$, $r$ and $i$,
straight segments of the stellar locus are centered at 0.  Bright,
unmasked stars ($r$<19) are selected and their GAaP magnitudes
corrected for Galactic extinction based on the
\cite{schlegel/etal:1998} maps and a standard $R_V = 3.1$ extinction
curve \footnote{At the relatively high Galactic latitudes where the KiDS 
survey areas are located the full dust path is probed at distances
between 0.5 and 1.0 kpc \citep{green/etal:2015}. Given the $r$-band 
brightness limit of $r\sim16$ this implies that 90\% of the stars
probed by KiDS are behind practically all the obscuring dust.}.
Per tile the offsets of the straight sections of the stellar
locus from 0 in each principal colour are minimized, and the zeropoint
offsets converted to three colour offsets: d($u-r$), d($g-r$) and
d($r-i$).

Comparing the colors of stars with SDSS measurements provides a
straightforward method to assess the improvement in the
calibration. In the following, the average color offsets found between
tiles in KiDS-North and SDSS DR9 are used. In $g-r$ the mean per-tile
difference between KiDS and SDSS is 6 mmag before SLR and 9 mmag
after, but the scatter (standard deviation) decreases from 38 mmag to
12 mmag. Similarly, in $r-i$ the mean offset and scatter change from
16 and 56 mmag to 6 and 11 mmag. Thus, while the absolute color
difference with SDSS remains comparable or improves slightly, the SLR
is very successful in homogenizing the color calibration. For $u-r$,
however, this is not the case, because both the mean color difference
and the scatter are significantly degraded, from 4 and 42 mmag to 80
and 64 mmag. For this reason the results for $u-r$ were not used to
calibrate the $u$-band zeropoints in the DR3 multi-band catalogue, where
the $u$-band thus purely relies on OP. On the other hand, in the 
KiDS-450 weak lensing shear catalogue (Sect. \ref{sec:wlcatalogs}) the
SLR results were applied in $u$-band, since no absolute calibration via
OP was used in that case. More details of the SLR procedure can
be found in \citet[Appendix B]{hildebrandt/etal:2017}.

\subsubsection{Accuracy and stability}

\begin{figure*}[t]
\centering
\includegraphics[width=\textwidth]{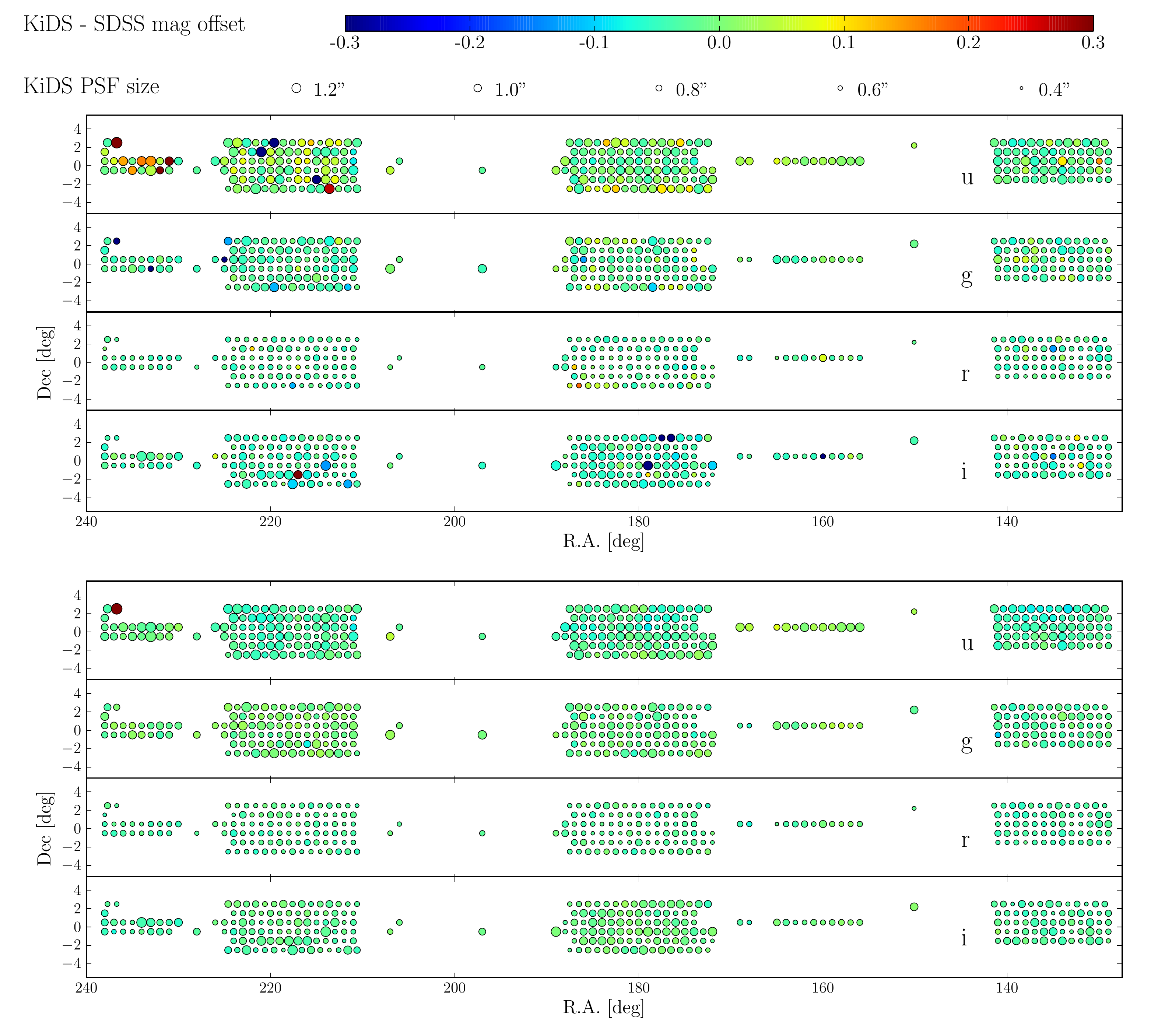}
\caption{Comparison of KiDS GAaP photometry before and after
  homogenization to SDSS DR13 for all tiles in KiDS-North as function
  of right ascension and declination. 
  {\it Top:} the magnitude offsets with respect to
  SDSS before applying photometric homogenization are indicated by the
  color scaling. The size of the circles represent the average PSF
  size in each coadded image. From top to bottom the panels correspond
  to the $u$, $g$, $r$, and $i$ filters. {\it Bottom:} same as top
  panel, but after applying photometric homogenization.}
\label{Fig:GaapVsSdss}
\end{figure*}

\begin{figure*}[t]
\centering
\includegraphics[width=\textwidth]{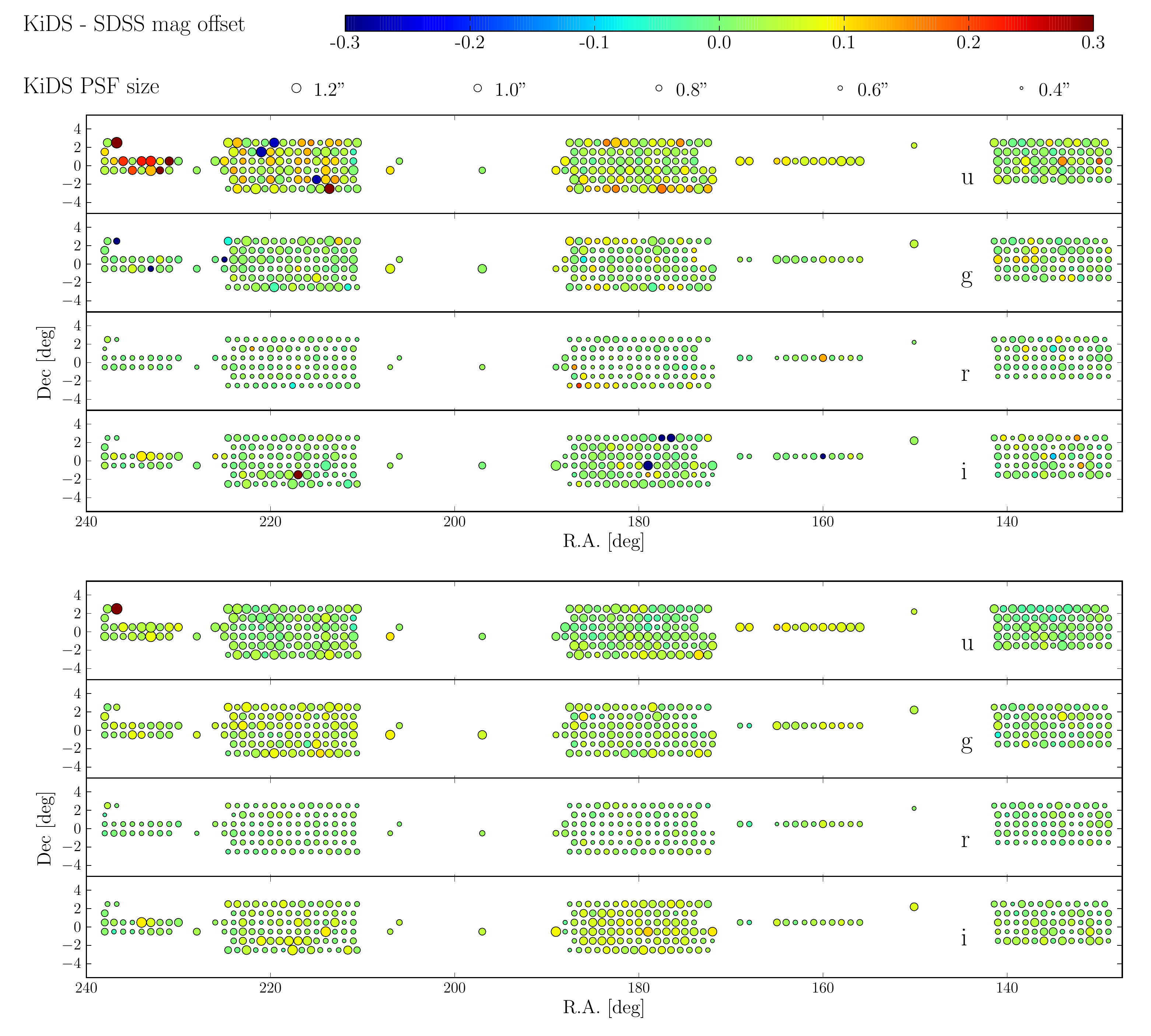}
\caption{Comparison of KiDS 10\arcsec~aperture-corrected 
  photometry before and after homogenization to SDSS DR13 for
  all tiles in KiDS-North as function of right ascension and
  declination. {\it Top:} the magnitude offsets with respect to
  SDSS before applying photometric homogenization are indicated by the
  color scaling. The size of the circles represent the average PSF
  size in each coadded image. From top to bottom the panels correspond
  to the $u$, $g$, $r$, and $i$ filters. {\it Bottom:} same as top
  panel, but after applying photometric homogenization.}
\label{Fig:Ap10VsSdss}
\end{figure*}

\begin{table*}
\caption{Comparison of KiDS-North and SDSS DR13 stellar photometry}
\label{Tab:KidsVsSdss}
\centering
\footnotesize
\begin{tabular}{l | c c | c c | c c | c c}
\hline\hline
Quantity & \multicolumn{4}{c|}{Not homogenized} & \multicolumn{4}{c}{Homogenized} \\
~ & \multicolumn{2}{c|}{GAaP} & \multicolumn{2}{c|}{10\arcsec~ap.cor} & \multicolumn{2}{c|}{GAaP} & \multicolumn{2}{c}{10\arcsec~ap.cor} \\
~ & Mean & $\sigma$ & Mean & $\sigma$ & Mean & $\sigma$ & Mean & $\sigma$\\
\hline
$u_\mathrm{KiDS}-u_\mathrm{SDSS}$ & 0.004 & 0.074 & 0.053 & 0.075 & $-$0.029 & 0.034 & 0.020 & 0.037 \\
$g_\mathrm{KiDS}-g_\mathrm{SDSS}$ & $-$0.028 & 0.074 & 0.014 & 0.074 & $-$0.011 & 0.023 & 0.031 & 0.028 \\
$r_\mathrm{KiDS}-r_\mathrm{SDSS}$ & $-$0.025 & 0.030 & 0.013 & 0.029 & $-$0.028 & 0.014 & 0.009 & 0.017 \\
$i_\mathrm{KiDS}-i_\mathrm{SDSS}$ & $-$0.041 & 0.057 & 0.011 & 0.055 & $-$0.020 & 0.017 & 0.032 & 0.028 \\
\hline
$(u-g)_\mathrm{KiDS}-(u-g)_\mathrm{SDSS}$ & 0.032 & 0.112 & 0.039 & 0.114 & $-$0.018 & 0.038 & $-$0.011 & 0.042 \\
$(g-r)_\mathrm{KiDS}-(g-r)_\mathrm{SDSS}$ & $-$0.003 & 0.078 & 0.002 & 0.077 & 0.017 & 0.015 & 0.022 & 0.028 \\
$(r-i)_\mathrm{KiDS}-(r-i)_\mathrm{SDSS}$ & 0.016 & 0.065 & 0.002 & 0.063 & $-$0.009 & 0.010 & $-$0.023 & 0.028 \\
\hline
\end{tabular}
\end{table*}

The quality of the final photometric homogenization, based on the
combination OP for the $u$- and $r$-band zeropoint corrections and SLR
for the $g$- and $i$-band corrections, can be quantified by a direct
photometric comparison of KiDS-North to measurements from SDSS. 
For the comparison presented in this section we use SDSS DR13 
\citep{albareti/etal:2016}, which includes a new 
photometric calibration \citep{finkbeiner/etal:2016} derived from
PanSTARRS DR1 \citep{chambers/etal:2016}, the most stable SDSS 
calibration to date. For this purpose we use stars that are
brighter than $r$ = 20, not flagged or masked in KiDS, and that have
photometric uncertainties smaller than 0.02 mag in SDSS and the
KiDS aperture magnitude in $g,r,i$ and 0.03 in $u$. Both
GAaP and 10\arcsec~aperture-corrected magnitudes are compared to the 
SDSS PsfMag magnitudes.

Figure \ref{Fig:HomVsSdss} shows the per-tile zeropoint offsets in each filter
as function of the per-tile difference between SDSS and the
uncorrected KiDS photometry. The clear correlation between these
quantities confirms that the photometric homogenization strategy works
as expected. Furthermore, this correlation is tighter for the GAaP
magnitudes than for the aperture-corrected magnitudes, which is not
surprising since the zeropoint offsets are derived based on the GAaP
results. 

The average per-tile photometric offsets between the KiDS DR3 tiles in
KiDS-North and SDSS, before applying the homogenization zeropoint
offsets, are illustrated in the top panels of Figures
\ref{Fig:GaapVsSdss} and \ref{Fig:Ap10VsSdss} for the GAaP and
10\arcsec~aperture-corrected photometry, respectively. In Table
\ref{Tab:KidsVsSdss} the mean offsets as well as the scatter are
listed. In all filters the scatter in the per-tile photometric offsets
is typically 5\%, although outliers, with offsets of several tenths of
a magnitude in some cases, are present in all filters. The result of
applying the zeropoint offsets discussed above are shown in the bottom
panels of Figures \ref{Fig:GaapVsSdss} and \ref{Fig:Ap10VsSdss}, and
the statistics again listed in Table \ref{Tab:KidsVsSdss}. Outliers
are successfully corrected by the procedure, both in the case of the
GAaP and the aperture-corrected photometry. The overall scatter in the
per-tile offsets is reduced with a factor 2 or more for GAaP, and up
to a factor 2 for the aperture magnitudes, clearly demonstrating the
improvement in the homogeneity of the calibration. Both before and
after homogenization systematic offsets of order 2\% between KiDS and
SDSS are visible that can be attributed to a combination of the
details of the absolute calibration strategy and colour terms between 
the KiDS and SDSS filters \citep[e.g.][]{dejong/etal:2015}, and the 
choice of photometric anchors for the OP. There is no sign of 
significant changes in the quality of the photometry between the 
different baffle configurations.

Since the SLR explicitly calibrates the $g$, $r$ and $i$ photometry
with respect to each other, the colour calibration between these
filters should be very good by definition. This is reflected in the
last two rows in Table \ref{Tab:KidsVsSdss}, where the $g-r$ and $r-i$
colours are compared to SDSS. In the case of GAaP the standard
deviation in the per-tile offsets is of order 1\% after
homogenization. Since the $u$ homogenization is independent from the
other filters, larger scatter remains in $u-g$, although also here
there is a large improvement in the stability of the colour
calibration.

\subsection{Photometric comparison to Gaia DR1}
\label{sec:gaia}

\begin{figure}
\centering
\includegraphics[width=\columnwidth]{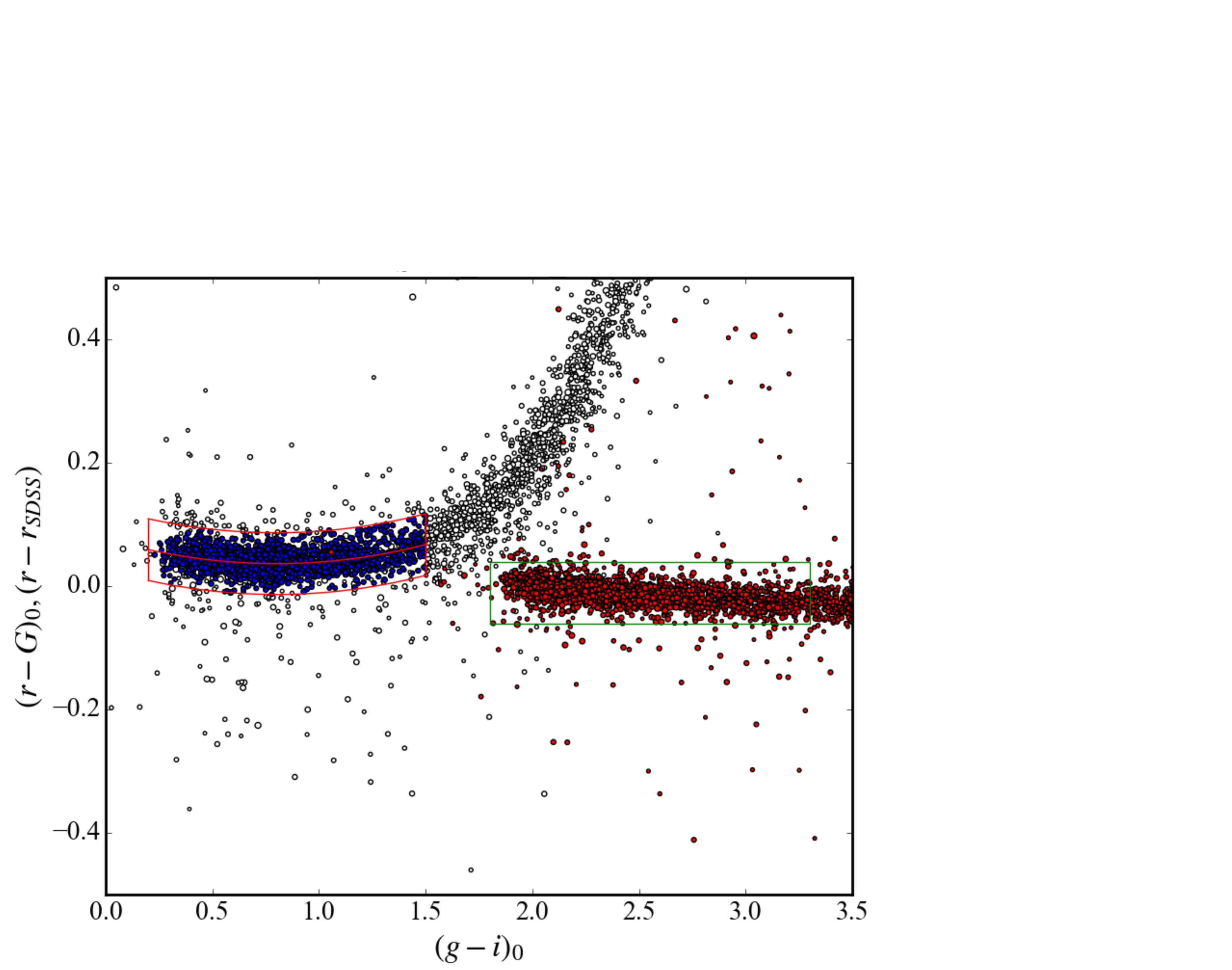}
\caption{Comparison of KiDS $r$-band GAaP photometry to Gaia DR1 $G$-band and SDSS DR9 $r$-band photometry. Stars with dereddened $g-i$ colours, based on colour-calibrated KiDS GAaP data ($x$-axis), between 0.2 and 1.5 are selected to iteratively derive the median photometric offsets $(r-G)_0$ (left sequence) and $(r-r_{\rm SDSS})$ (right sequence). The latter sequence of data points is shifted by 1.6 mag in $(g-i)_0$ for clarity. The red and green outlines show the regions encompassing the data points used in the final iteration of each fit.}
\label{Fig:KidsGaiaSdssZpt}
\end{figure}

The first data release of Gaia \citep[Gaia DR1,][]{gaia/etal:2016}
provides a photometrically stable and accurate catalog
\citep{vanleeuwen/etal:2016} to which both the KiDS-North and
KiDS-South fields can be compared, thus allowing a validation of the
photometric homogeneity over the full survey. The photometry released
in Gaia DR1 is based on one very broad `white light' filter, $G$, that
encompasses the KiDS $gri$ filters \citep{jordi/etal:2010}, allowing
photometric comparisons. For this purpose, the DR3 multi-band catalog
was matched to the Gaia DR1 catalog, as well as the SDSS DR9
photometric catalogue, yielding between 1\,000 and 5\,000 matched
point-like sources per KiDS survey tile.  Since the $r$-band data is
the highest quality in KiDS and also used for the absolute
calibration, this was also our choice for a direct comparison to the
$G$-band. Although the central wavelengths of KiDS $r$-band and Gaia
$G$-band are similar, the difference in wavelength coverage does lead
to a difference in the attenuation due to foreground extinction. To
assess this, quadratic fits of the stellar locus in $G_0 - r_0$ vs $(g
- i)_0$ were performed, using extinction corrected SDSS photometry for
stars with $0.6 < (g - i)_0 < 1.6$. Solving for the extinction in $G$
in four iterations, we find a relative attenuation $A_G/A_r$ = 0.93
$\pm$ 0.01, meaning that the extinction in $G$ is very close to that
in $r$-band. Considering the small reddening in the KiDS fields, which
is at most $E(B-V) \sim 0.1$ but much smaller almost everywhere in
the survey area, the effect of this difference is always less than
0.01.

Subsequently we move to using KiDS data, comparing the $r$-band GAaP
to both Gaia $G$ measurements and to SDSS $r$-band PSF
magnitudes. Stars with $0.2 < (g-i)_{\rm GAaP} < 1.5$, color-calibrated
and extinction corrected, are selected and the median offset between
$r_{\rm GAaP}$ and $G$ and between $r_{\rm GAaP}$ and $r_{\rm SDSS}$ are
iteratively determined in a box of width $\pm$0.05 mag, as illustrated
in Figure \ref{Fig:KidsGaiaSdssZpt}. Extinction corrections are
applied for each star, based on \cite{schlegel/etal:1998} and the
value of $A_G$ derived above. We consistently find an offset between
$(r_{\rm GAaP}-G)$ and $r_{\rm GAaP}-r_{\rm SDSS}$ of 0.049 with a scatter of only
0.01 mag. This allows an indirect derivation via the Gaia data of the
offset between the GAaP $r$-band photometry and SDSS even in the
absence of SDSS data, such as in the KiDS-S field.

\begin{figure*}[t]
\centering
\includegraphics[width=\textwidth]{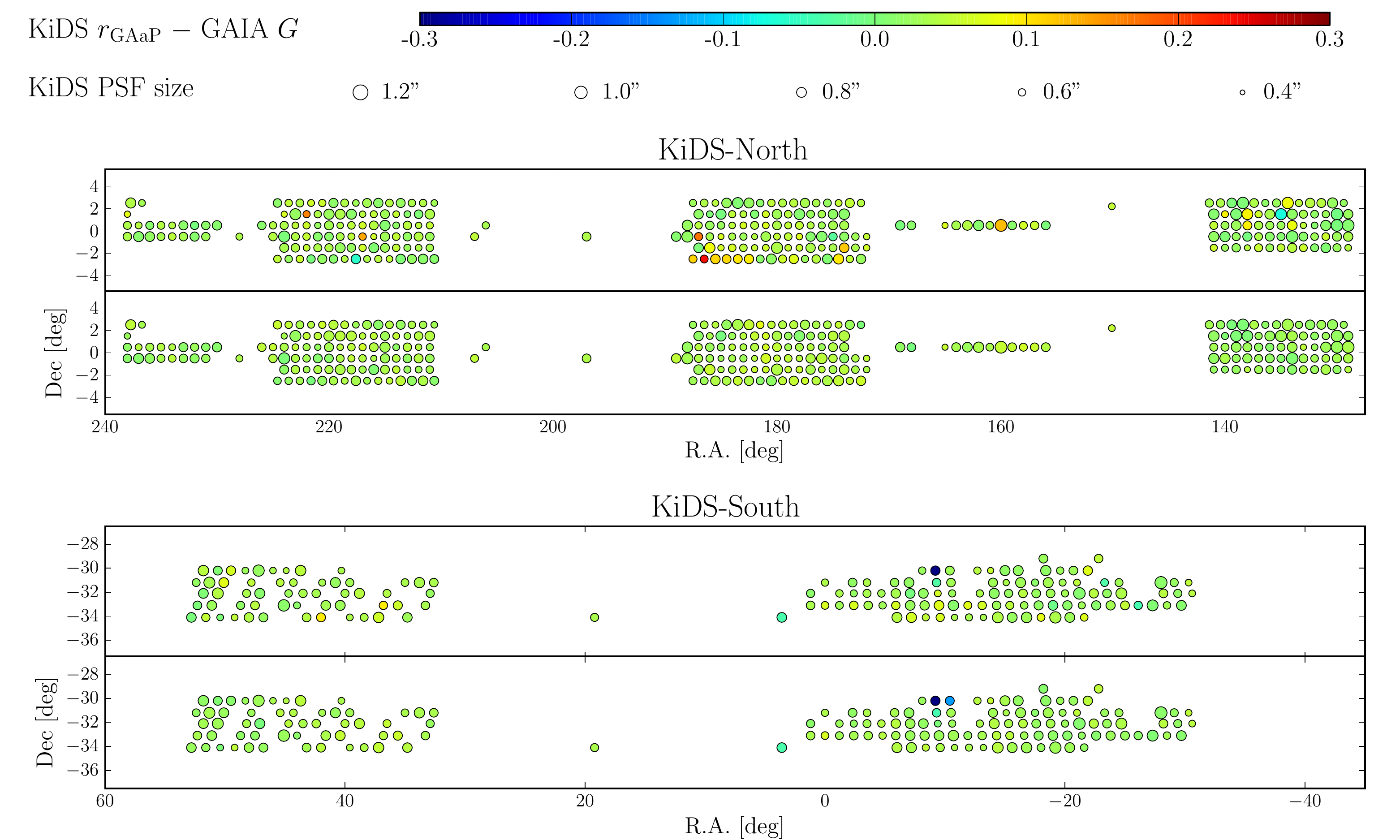}
\caption{Comparison of KiDS $r$-band to Gaia $G$-band photometry as
  function of position on the sky. The per-tile median photometric
  offset ($r_{\rm GAaP}-G$) is indicated by the color scale, while the marker size
  corresponds to the mean seeing in the KiDS $r$-band image. {\it
    Top:} Comparison for KiDS-North, with in the top and bottom
  subpanels showing the offsets before and after applying the
  photometric homogenization, respectively. {\it Bottom:} same as the
  top panel, but now for KiDS-South.}
\label{Fig:KiDSvsGaia}
\end{figure*}

Following this strategy, the absolute calibration of the GAaP $r$-band
photometry has been verified for all tiles in KiDS-ESO-DR3. Figure
\ref{Fig:KiDSvsGaia} shows the photometric offsets between $r_{\rm GAaP} -
G$ for the full data set both before and after applying the
photometric homogenization based on overlap photometry. Table
\ref{Tab:KidsVsGaia} summarizes the statistics of this comparison. For
KiDS-North the overall picture is the same as for the direct
comparison to SDSS (Fig. \ref{Fig:GaapVsSdss}). Also in KiDS-South the
homogenization works as expected, although one weakness of the current
strategy is clearly revealed. The tile KIDS\_350.8\_-30.2 turns out to
show a large photometric offset in $r$-band, but unfortunately was
selected as a photometric anchor according to the criteria listed in
Sect. \ref{sec:phothom}. As a result this offset persists after the
homogenization, and also has a detrimental effect on a neighbouring
tile. This is reflected in the standard deviation of the offsets in
KiDS-South, which is not significantly improved by the homogenization.
From this analysis, the average offset between KiDS $r$-band and SDSS
$r$-band is shown to be approximately $-0.015$, which can be largely
attributed to the colour term in $r_{\rm KiDS} - r_{\rm SDSS}$ that is
apparent from the tilt in the sequence of red points in
Fig. \ref{Fig:KidsGaiaSdssZpt}.

The comparison with the Gaia DR1 $G$-band photometry shows the
tremendous value of this all-sky, stable photometric catalogue for the
validation, and possibly calibration, of ground-based surveys such as
KiDS. Since KiDS-ESO-DR3 was released before these data became
available, they are only used as a validation for the photometric
calibration. However, in case of the shear catalogue described in
Sect. \ref{sec:wlcatalogs} we provide per-tile photometric offsets in
the catalogue itself that allow the photometry to be homogenized based
on the comparison with Gaia $G$ data.  We are currently studying the
possibilities for using Gaia data to further improve the photometric
calibration of the KiDS photometry for future data releases. 
Although the Gaia DR1 catalogue still contains areas that are too 
sparse to use for our astrometric calibration, we anticipate moving from
2MASS \citep{skrutskie/etal:2006} to Gaia as astrometric reference 
catalogue once Gaia DR2 becomes available.
 
\begin{table}
\caption{Comparison of $r$-band GAaP and Gaia DR1 photometry}
\label{Tab:KidsVsGaia}
\centering
\footnotesize
\begin{tabular}{l | c c c | c c c }
\hline\hline
Field & \multicolumn{3}{c|}{Not homogenized} & \multicolumn{3}{c}{Homogenized} \\
~ & $r-G$ & $r-r_{\rm SDSS}$ & $\sigma$ & $r-G$ & $r-r_{\rm SDSS}$ & $\sigma$ \\
\hline
\hline
KiDS-N & 0.036 & $-$0.013 & 0.030 & 0.032 & $-$0.017 & 0.014 \\
KiDS-S & 0.029 & $-$0.020 & 0.035 & 0.025 & $-$0.024 & 0.034 \\
Total & 0.033 & $-$0.016 & 0.032 & 0.030 & $-$0.019 & 0.023 \\
\hline
\end{tabular}
\end{table}

\section{Weak lensing shear data}
\label{sec:weaklensing}

For the weak gravitational analyses of KiDS accurate shear estimates
of small and faint galaxy images are measured from the $r$-band
data. This imposes especially strict requirements on the quality of
the astrometric calibration \citep{miller/etal:2013}. Furthermore, because weak lensing
measurements are intrinsically noise-dominated and rely on ensemble
averaging, small systematic shape residuals can significantly affect
the final results. For this reason, shears are measured based on a 
joint fit to single
exposures rather than on image stacks, avoiding any systematics
introduced by re-sampling and stacking of the image pixels.
Therefore, a dedicated pipeline that
has already been successfully used for weak lensing analyses
in previous major Wide-Field-Imaging surveys
\citep[e.g.][]{heymans/etal:2012,erben/etal:2013,hildebrandt/etal:2016}
is employed to obtain optimal shape measurements from the $r$-band data.
This dedicated pipeline makes use of
\textsc{THELI} \citep{erben/etal:2005,schirmer:2013} and the
\textit{lens}fit shear measurement code \citep{miller/etal:2013,fenechconti/etal:2016}. In
the following subsections, the additional pixel processing and the
creation of the weak lensing shear catalogue are reviewed.

\subsection{Image data for weak lensing}
\label{sec:theli}

The additional $r$-band data reduction was done with the
\textsc{THELI} pipeline \citep{erben/etal:2005,schirmer:2013}. A
detailed description of our prescription to process OmegaCAM data and
a careful evaluation of the data quality will be provided in a
forthcoming publication (Erben et al., in preparation). We therefore
only give a very short description of essential processing steps:
\begin{enumerate}
  \item The initial data set for the \textsc{THELI} processing is identical to
    that of DR3 and consists of all $r$-band data observed between the
    9th of August 2011 and the 4th of October 2015. The raw data is
    retrieved from the ESO archive (see \url{http://archive.eso.org}).
      \item Science data are corrected for crosstalk effects. We measure
    significant crosstalk between CCDs \#94, \#95 and \#96. Each pair
    of these three CCDs show positive or negative crosstalk in both
    directions. We found that the strength of the flux transfer
    varies on short time-scales and we therefore
    determine new crosstalk coefficients for each KiDS observing
    block. An $r$-band KiDS observing block extends to about 1800 s
    in five consecutive exposures.
      \item The removal of the instrumental signature (bias subtraction,
    flat-fielding) is performed
    simultaneously on all data from a two-weeks period around each
    new-moon and full-moon phase. Hence, two-week periods of
    moon-phases define our \textit{runs} (\textit{processing run} in
    the following) for OmegaCAM data processing 
    \citep[see also Sect. 4 of][]{erben/etal:2005}.  We tested that the instrument
    configuration is stable for a particular processing run. The
    distinction of runs with moon phases also naturally reflects usage
    of certain filter combinations on the telescope ($u$, $g$
    and $r$ during new-moon and $i$ during full-moon phases).
      \item Photometric zeropoints are estimated for all data of a complete
    processing run. They are obtained by calibration of \textit{all}
    science images in a run which overlap with the Data Release 12 of
    the SDSS \citep[see][]{alam/etal:2015}. We have between 30 and 150 such
    images with good airmass coverage for each processing run.
    We note that we did not carry out extensive tests on the quality of our
    photometric calibration. As all required photometric analysis
    within weak lensing projects is performed with the DR3 data set, a
    rough photometric calibration of the THELI data is sufficient for
    our purposes. 
      \item As a last step of the \textit{run} processing we subtract the
    sky from all individual chips. These data form the basis for the later
    object shape analysis -- see \cite{hildebrandt/etal:2017} and
    Sect.~\ref{sec:wlcatalogs}.
      \item Since the astrometric calibration is particularly crucial for
    an accurate shear estimate of small and faint galaxy images, great
    care was taken in this part of the analysis
    and we used all available information to obtain optimal
    results. The available KiDS tiles were divided into five patches, three in KiDS-North and two in KiDS-South, centered around large contiguous areas. We simultaneously astrometrically calibrate \emph{all}
    data from a given patch, i.e., we perform a patch-wide global
    astrometric calibration of the data. This allows us to take into
    account information from overlap areas of individual KiDS
    tiles. For the northern patches and their southern
    counterparts our processing differed depending on the availability of
    additional external data:
        \begin{enumerate}
      \item For the northern KiDS patches we use accurate astrometric
      reference sources from the SDSS-Data Release 12
      \citep{alam/etal:2015} for the absolute astrometric reference
      frame. No additional external data were available here.
            \item The southern KiDS-patches do not overlap with the
      SDSS, and we have to use the less accurate 2MASS catalogue (see
      \citealt{skrutskie/etal:2006}) for the absolute astrometric
      reference frame. However, the area of these patches is covered
      by the public VST ATLAS Survey \citep{shanks/etal:2015}. ATLAS
      is significantly shallower than KiDS (each ATLAS pointing
      consists of two 
            $45s$
      OmegaCAM exposures) but it
      covers the area with a different pointing footprint than
      KiDS. This allows us to constrain optical distortions better,
      and to compensate for the less accurate astrometric 2MASS
      catalogue. Our global patch-wide astrometric calibration
      includes \emph{all} KiDS and ATLAS $r$-band images covering the
      corresponding area.
    \end{enumerate}
  \item The astrometrically calibrated data are co-added with a
    weighted mean approach \citep[see][]{erben/etal:2005}.  The
    identification of pixels that should not contribute and pixel
    weighting of usable regions is done as described in
    \citet{erben/etal:2009,erben/etal:2013} for data from the
    Canada-France-Hawaii Telescope Legacy Survey.  The set of images entering
    the co-addition is identical to the set used for the KiDS DR1, DR2 and DR3 
    coadds. The final products of the THELI processing are the
    single-chip data (used for shear measurements, see above), the 
    co-added science image (used for source detection), a
    corresponding weight map and a \textit{sum} image
    \citep[for a more detailed description of these
    products see also][]{erben/etal:2013}.
\end{enumerate}

\subsection{Weak lensing catalogue}
\label{sec:wlcatalogs}

In addition to the KiDS-ESO-DR3 catalogues we also release the
catalogue that was used for the KiDS-450 cosmic shear project
\citep{hildebrandt/etal:2017}. This catalogue differs in some aspects
from the DR3 catalogues as described in the following:
\begin{itemize}
  \item The KiDS $r$-band data have been reduced with an independent
    data reduction pipeline, THELI \citep{erben/etal:2005} that is
    optimised for weak lensing applications. See
    Sect. \ref{sec:theli}, as well as \citet{hildebrandt/etal:2017} and \citet{kuijken/etal:2015} for details.
  \item Source detection is performed on stacks reduced with THELI (see Sect.\ref{sec:theli}). Hence, the source lists are slightly different from the DR3 catalogues.
  \item Multi-colour photometry and photo-z are estimated in the
    same way as for DR3 (Sect. \ref{sec:photz}) except that the source list is based on the THELI stacks.
  \item The \textit{lens}fit shear measurement code \citep{miller/etal:2013} is run on the individual exposures calibrated by THELI. This yields accurate ellipticity measurements and associated weights for each source.
  \item Image simulations are used to calibrate the shear measurements as described in \citet{fenechconti/etal:2016}. A multiplicative shear bias correction based on these results is included for each galaxy.
  \item Masks for the weak lensing catalogue include defects detected
    on the THELI stacks and differ slightly from the DR3 masks. This
    masking information is also included in the lensing catalogues.
  \item Contrary to the general purpose DR3 multi-band catalogue, the photometry is colour-calibrated, but the absolute
    calibration is not homogenized. Stellar locus regression was used to colour calibrate the $u$, $g$ and $i$ filters to the $r$-band, while no overlap photometry scheme was applied.
    In the catalogue now publicly available zeropoint offsets derived from
    Gaia DR1 photometry (Sect. \ref{sec:gaia}) are included in the catalogue
    and can be used to homogenize the absolute calibration.
\end{itemize} 

More details can be found in \cite{hildebrandt/etal:2017}. 
Descriptions of all columns in this weak lensing catalogue can be found in appendix~\ref{App:shearcat} and Table~\ref{Tab:ShearCatColumns}.

\section{Photometric redshifts}
\label{sec:photz}

Several sets of photometric redshifts are publicly available, based on
the KiDS DR3 data set. These include photo-z's derived using the BPZ
template fitting method \citep{benitez:2000}, as well as photo-z's
based on two different machine-learning techniques. In the following
sections we discuss the various sets of photo-z's, followed by a
discussion in Sect. \ref{sec:photz-comparison} of their performance and
applicability to specific science cases.

The following sections describe the computation of the different 
sets of photo-z's. In Sect. \ref{sec:photz-bpz} the two available BPZ
data sets are discussed and compared, while Sect. \ref{sec:photz-mlpqna}
and \ref{sec:photz-annz2} focus on the two machine-learning based data
sets. Finally, in Sect. \ref{sec:photz-comparison} all photo-z sets are
juxtaposed with the same spectroscopic data sets in order to provide an
objective comparison between them, together with recommendations for
different use cases. In all places where photometric redshifts are 
compared to spectroscopic redshifts, relative errors in the photo-z's 
are defined as
\begin{equation}
\delta z = (z_{\rm phot} - z_{\rm spec})/(1 + z_{\rm spec})
\label{eq:deltaz}
\end{equation}
and catastrophic outliers as having $|\delta z| > 0.15$. 
Furthermore, $\sigma$ and NMAD denote the standard deviation and the
normalized median absolute deviation, the latter defined as
\begin{equation}
{\rm NMAD} = 1.4826 \, {\rm median}(|\delta z - {\rm median}(\delta z)|),
\label{eq:nmad}
\end{equation}
and reported values are always without clipping of outliers.

\subsection{BPZ photometric redshifts}
\label{sec:photz-bpz}

\begin{figure}
   \centering
   \includegraphics[width=\columnwidth]{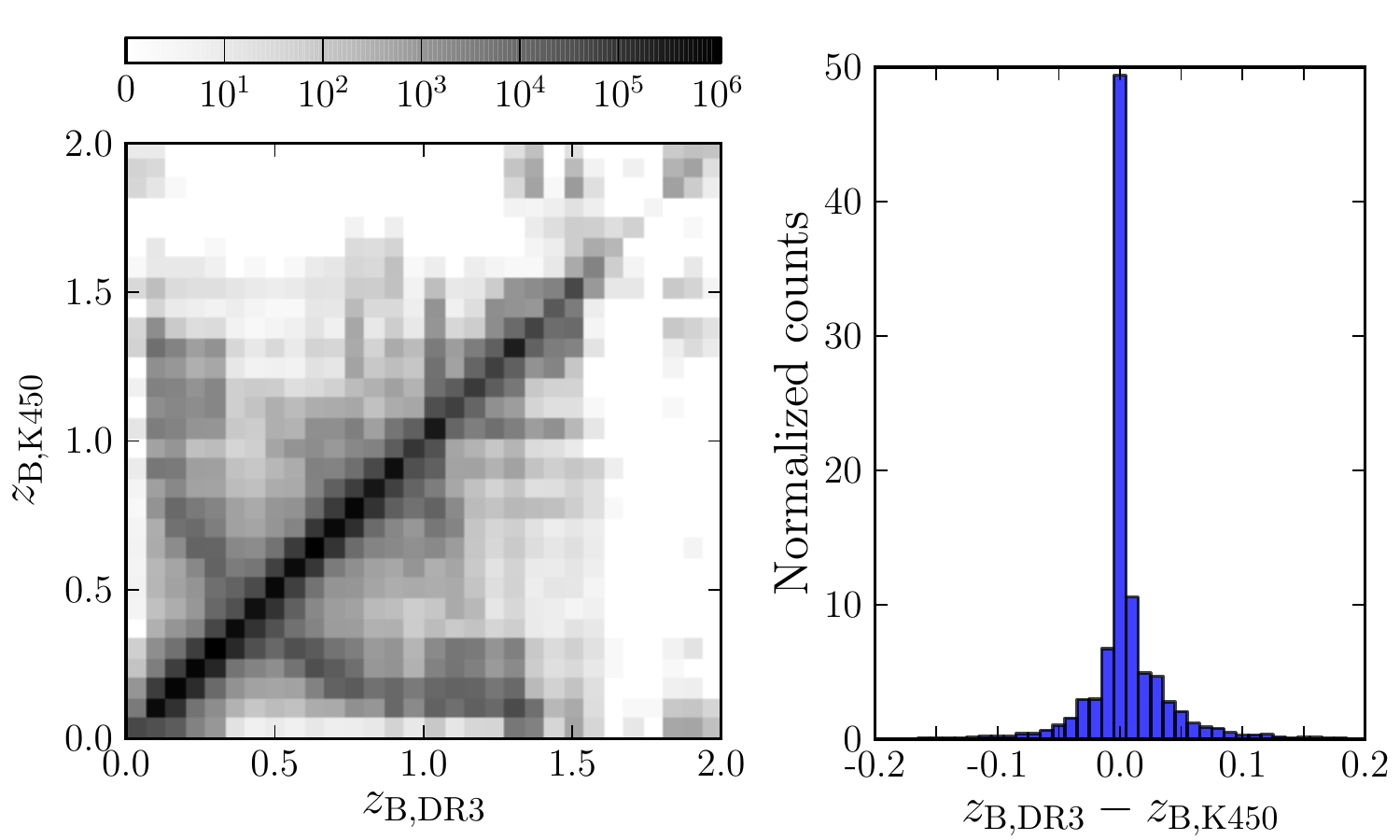}
   \caption{Comparison of the BPZ photo-z's included in the DR3
     multi-band catalogue ($z_{\rm B,DR3}$) and in the KiDS-450 shear
     catalogue ($z_{\rm B,K450}$). {\it Left:} 2D histogram of the
     direct comparison. The grey scale indicating the number of
     sources per bin is logarithmic. {\it Right:} normalized
     histogram of $z_{\rm B,DR3} - z_{\rm B,K450}$.}
   \label{Fig:DR3vsK450BPZ}
\end{figure}

Both the KiDS DR3 multi-band catalogue (Sect. \ref{sec:dr3}), as well
as the KiDS-450 weak lensing shear catalogue
(Sect. \ref{sec:wlcatalogs}), include photometric redshifts based on
the Bayesian template fitting photo-z code pioneered by
\cite{benitez:2000}.  These photo-z's are calculated following the
methods developed for CFHTLenS
\citep{heymans/etal:2012,hildebrandt/etal:2012}, making use of the
re-calibrated templates from \cite{capak:2004}. A more detailed
description of the procedures applied specifically to the KiDS data
can be found in \cite{kuijken/etal:2015}.  Included in the DR3 catalog
(see Table \ref{Tab:MultiBandColumns}) are
the best-fit photometric redshift $z_{\rm B}$ and the best-fit
spectral type. Since the spectral types are interpolated, the best-fit
value does not necessarily correspond to one type, but can be
fractional. The catalog also includes the ODDS parameter, which is a
measure of the uni-modality of the redshift probability distribution
function (PDF) and can be used as a quality indicator. Not included in
the catalog are the posterior redshift PDFs, which are available for
download from the DR3
website\footnote{\url{http://kids.strw.leidenuniv.nl/DR3/bpz-pdfs.php}}.
The KiDS-450 shear catalogue also provides $z_{\rm B}$, ODDS and the
best-fit template, but additionally also the lower and upper bounds of
the 95\% confidence interval of $z_{\rm B}$ (see Table
\ref{Tab:ShearCatColumns}). 

Since photo-z's critically rely on accurate colours, both sets of BPZ
photo-z's make use of GAaP photometry (Sect. \ref{sec:gaap}), measured
on the same \textsc{Astro-WISE}-reduced $ugri$ coadded images from the
KiDS public data releases. There are, however, some differences
between the two data sets. Source detection for the DR3 multi-band
catalogue is performed on the \textsc{Astro-WISE}-reduced $r$-band
coadds, while for the KiDS-450 shear catalogue this is done on the
\textsc{THELI}-reduced $r$-band coadds. Because the astrometry is
derived independently between these reductions, there exist small
(typically subpixel) offsets between the positions. The same small
offsets exist between the GAaP apertures used for the two data sets,
which can introduce small photometric differences.  Furthermore, for
KiDS-450 the GAaP $ugri$ photometry was colour-calibrated using
stellar locus regression, with no homogenization of the absolute
calibration. On the other hand, for DR3 only the $gri$ filters were
colour-calibrated using stellar locus regression, while the $u$- and
$r$-band absolute calibration were independently homogenized
(Sect. \ref{sec:phothom}).  Although the absolute calibration of the
photometry affects the BPZ priors, this is not expected to have a
major effect on the redshift estimates. There might be some small
effect from the difference in the calibration of the $u$-band between
the two data sets. Finally, the KiDS-450 shear catalogue is limited to
relatively faint sources with $20 < r < 25$.

A direct comparison between the two BPZ photo-z sets is shown in the
left panel of Figure \ref{Fig:DR3vsK450BPZ}. The vast majority of
sources lie along the diagonal, with almost mirror symmetric
low-level structure visible on either side. The normalized
histogram of the differences between the two sets of photo-z's (right
panel of Figure \ref{Fig:DR3vsK450BPZ}) shows a narrow peak centered
at 0.  The photo-z resolution is 0.01 and 49\% of the sources have the
same best photo-z estimate, $z_{\rm B}$, both in the DR3 and in the
KiDS-450 data set; 65\% of all sources have values of $z_{\rm B}$ that
agree within 0.01 and 93\% within 0.1. To verify the assumption 
that the lack of homogenization of the absolute $r$-band calibration in
KiDS-450 does not significantly affect the photo-z results, these same
comparisons were done limited to the tiles with derived zeropoint 
offsets larger than 0.1 mag. For this set of 15 tiles the results are
very similar, with 45\% of sources having identical photo-z and 64\% 
and 93\% agreeing to within 0.01 and 0.1, respectively, confirming that
the absolute calibration does not have a substantial effect.

There are no signs of
significant biases or trends between the two sets of BPZ photo-z's, and
we conclude that the two BPZ based sets of photo-z's can be 
considered to be consistent for most intents and purposes.

\subsection{MLPQNA photometric redshifts and Probability Distribution Functions}
\label{sec:photz-mlpqna}

Photometric redshifts for KiDS DR3 have also been produced using the
MLPQNA (Multi Layer Perceptron with Quasi Newton Algorithm) machine
learning technique, following the analysis for KiDS DR2
\citep{cavuoti/etal:2015}. This supervised neural network was already
successfully employed in several photometric redshift experiments
\citep{biviano/etal:2013,brescia/etal:2013,brescia/etal:2014}.
The typical mechanism for a supervised machine learning method to
predict photometric redshifts foresees the creation of a Knowledge
Base (hereafter KB), split into a training set for the model learning
phase and a blind test set for evaluating the overall performance of
the model. The term `supervised' implies that both training and test
sets must contain objects for which the spectroscopic redshift,
e.g. the ground truth, is provided.

In the specific case of DR3, the KB used for MLPQNA is composed of
$214$ tiles of KiDS data cross-matched with SDSS-III data release $9$
\citep{ahn/etal:2012} and GAMA data release $2$
\citep{liske/etal:2015} merged spectroscopic redshifts. The photometry
is based on the $ugri$ homogenized magnitudes, based on the GAaP
measurements (hereafter mag\_gaap bands), two aperture magnitudes,
measured within circular apertures of 4\arcsec and 6\arcsec 
diameter, respectively, corrected for extinction and zeropoint offsets, and
related colours, for a total of $21$ photometric parameters for each
object.
The initial combination of the tiles leads to $120\,047$ objects, after
which the tails of the magnitude distributions and sources with
missing magnitude measurements were removed.

\begin{figure*}
\centering
\includegraphics[width=\textwidth]{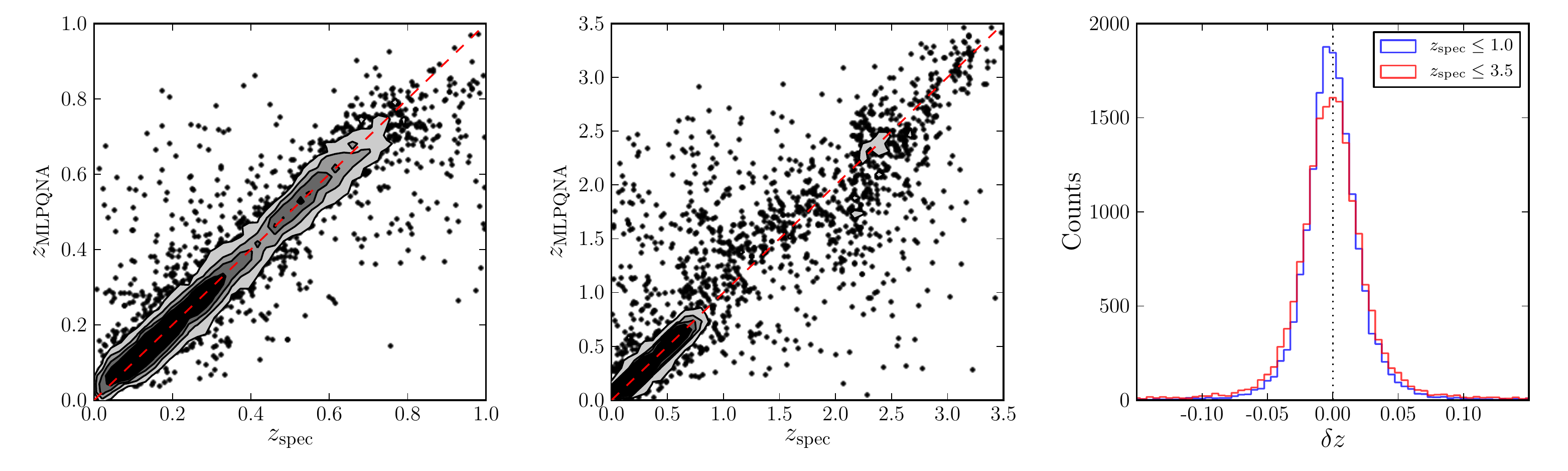}
\caption{Results of the MLPQNA photo-z experiments. {\it Left:} Scatter plot of $z_\mathrm{spec}$ vs $z_\mathrm{phot}$ for the experiment limiting
  $z_\mathrm{spec}$ between $0.01$ and $1$. {\it Centre:} Same as the left
  panel but for $z_\mathrm{spec}$ between $0.01$ and $3.5$. {\it
    Right:} Histograms of the residual distribution for the experiment
  with $0.01 < z_\mathrm{spec} < 1.0$ (blue) and $0.01 <
  z_\mathrm{spec} < 3.5$ (red). }
\label{Fig:MlpqnaTests}
\end{figure*}

Two separate experiments were performed, within different $z_\mathrm{spec}$
ranges: i) $0.01 \leq z_\mathrm{spec} \leq 1$ and ii) $0.01 \leq z_\mathrm{spec}
\leq 3.5$. After cleaning we subdivided
the data into training and test sets. For experiment i) we
obtained $66\,731$ objects for training and $16\,742$ test objects and for
experiment ii) $70\,688$ training objects and $17\,659$ test objects.
Scatter plots for the two experiments are shown,
respectively, in the left and middle panels of Figure \ref{Fig:MlpqnaTests}.

The statistics, calculated for the quantity $\delta z$
(Eq. \ref{eq:deltaz}), obtained for experiment i) are mean $\delta z$
= 0.0014, $\sigma$ = 0.035 and NMAD (Eq. \ref{eq:nmad}) = 0.018, with
$0.93\%$ outliers ($|\delta z| >0.15$). The distribution of residuals is
shown by the blue histogram in the right panel of
Fig.~\ref{Fig:MlpqnaTests}.
For experiment ii) the values are the following:
$\overline{\delta z}$ = 0.0063, $\sigma$ = 0.101 and NMAD = 0.022,
with $3.4\%$ outliers. In this case the residual distribution is shown
in red in the right panel of Fig.\ref{Fig:MlpqnaTests}.

The photometric redshifts are characterized by means of a
photo-z Probability Density Function (PDF), derived by the 
METAPHOR (Machine-learning Estimation Tool for Accurate PHOtometric
Redshifts) method, designed to provide a reliable PDF of the error
distribution for empirical models \citep{cavuoti/etal:2017}. The
method is implemented as a modular workflow, whose internal engine for
photo-z estimation makes use of the MLPQNA neural network.  The
METAPHOR pipeline is based on three functional macro phases:
\begin{itemize}
 \item Data Pre-processing: data preparation, photometric evaluation
   and error estimation of the KB, followed by its perturbation;
 \item Photo-z prediction: training/test phase performed through the
   MLPQNA model; 
 \item PDF estimation: production of the photo-z's PDF and evaluation
   of the statistical performance. 
\end{itemize}

Given the spectroscopic sample, randomly shuffled and split into
training and test sets, we proceed by training the MLPQNA model and by
perturbing the photometry of the given test set to obtain an arbitrary
number $N$ test sets with a variable photometric noise
contamination.  Then we submit the $N+1$ test sets (i.e. $N$ perturbed
sets plus the original one) to the trained model, thus
obtaining $N+1$ estimates of photo-z. With these $N+1$ values we
perform a binning in photo-z, thus calculating for each one the
probability that a given photo-z value belongs to each bin. We
selected a binning step of $0.01$ for the described experiments.

The photometry perturbation can be selected among a series of types,
described in \cite{cavuoti/etal:2017}. Here the choice is based on the
following expression, which is applied to the given \textit{j}
magnitudes of each band \textit{i} as many times as the number of
perturbations of the test set:
\begin{equation}\label{eq1}
m_{ij, \textrm{perturbed}}=m_{ij}+\alpha_{i}F_{ij} \, u_{(\mu=0,\sigma=1)}
\end{equation}
where $\alpha_i$ is a multiplicative constant, defined by the user
and $F_{ij}$ is a band-specific bimodal function 
designed to realistically scale the contribution of Gaussian noise
to the magnitudes. To this end, for each magnitude $F_{ij}$ takes the 
maximum of a constant, heuristically chosen for each band, or a 
polynomial fit to the magnitude-binned photometric errors; in other 
words, at the faint end $F_{ij}$ follows the derived photometric 
uncertainties and at the bright end it has a user-defined constant 
value. Finally, the term $u_{(\mu=0,\sigma=1)}$ in
Eq. \ref{eq1} is randomly drawn from a Gaussian centered on 0 
with unit variance. A detailed description of the produced photo-z 
catalogue will be provided in Amaro et al. (in prep.).

The final photo-z catalogue produced consists of 8\,586\,152 objects, 
by including all data compliant with the magnitude ranges imposed by 
the KB used to train our model and specified in Appendix 
\ref{App:mlphotz}. For convenience, the whole catalogue was split into 
two categories of files, namely a single catalogue file with the best
predicted redshifts for the KiDS DR3 multi-band catalogue, and a set of 
440 files, one for each included survey tile, that contain the photo-z 
PDFs. The file formats are specified in Tables \ref{Tab:mlpqna-photz} 
and \ref{Tab:mlpqna-pdf}.

\subsection{ANNz2 photometric redshifts}
\label{sec:photz-annz2}

Photometric redshifts based on another machine learning method, namely 
ANNz2 \citep{sadeh/etal:2016}, are provided for KiDS DR3 as well. This
versatile tool combines various ML approaches (artificial neural
networks, boosted decision trees, etc.) and allows the user to derive
photometric redshifts and their PDFs based on spectroscopic training
sets. It also incorporates a weighting scheme, using a kd-tree
algorithm, which enables weighting the training set to mimic the
photometric properties of the target data (see
e.g. \citealt{lima/etal:2008}). A similar principle has been applied
in the KiDS cosmic shear analysis by \cite{hildebrandt/etal:2017} to
directly calibrate photometric redshift distributions based on deep
spectroscopic data. Here we use a similar set of redshift samples as
in \cite{hildebrandt/etal:2017} for the training, including fields outside of the main KiDS
footprint, which were purposely observed for the survey, namely: a
non-public extension of zCOSMOS data \citep{lilly/etal:2009}, kindly
shared by the zCOSMOS team; an
ESO-released\footnote{\url{http://www.eso.org/sci/activities/garching/projects/goods/MasterSpectroscopy.html}} compilation of spec-z's in the
CDFS field; as well as redshift data from two DEEP2 fields
\citep{newman/etal:2013}. This is supplemented by redshift samples not
used by \cite{hildebrandt/etal:2017}, i.e. GAMA-II data\footnote{Note
  that the GAMA-II data are deeper and more complete than the GAMA DR2
data used in Sect. \ref{sec:photz-mlpqna} and \ref{sec:photz-comparison}}
\citep{liske/etal:2015}; 2dFLenS measurements \citep{blake/etal:2016};
SDSS-DR13 galaxy spectroscopy \citep{sdss:2016}; 
additional redshifts
in the COSMOS field from the "GAMA G10" analysis
\citep{davies/etal:2015}; and extra redshifts in the CDFS field from
the ACES survey \citep{cooper/etal:2012}. After combining all these
datasets and removing duplicates we have over $300\,000$ sources
detected also by KiDS that are used for training and tests of the
photometric redshift accuracy. We note however that the bulk of these
($\sim200\,000$) come from GAMA which provides information only at
$r<19.8$. It is the availability of the deep spectroscopic data
covered by KiDS imaging that allows us to derive machine-learning
photo-z's at almost the full depth of the photometric sample. The details of
the methodology and photo-z statistics will be provided in Bilicki et
al. (in prep.). Here we report the main results of our experiments,
and compare them with the fiducial KiDS photo-z's available from BPZ.

The full KiDS spectroscopic sample is currently dominated by
relatively low-redshift ($\overline{z} = 0.314$) and bright
galaxies. By splitting it randomly
into training and test sets we were thus able to test the performance
of ANNz2 for KiDS in this regime. The experiments indicate better
results for ANNz2 than for BPZ at these low redshifts: mean bias of
$\delta z$ equal to 0.0015, with $\sigma$ = 0.068 and NMAD = 0.025 and 3.3\% outliers as compared to
$\overline{\delta z} = 0.0007$, $\sigma = 0.086$, NMAD $= 0.035$ and
3.9\% outliers for the same sources from BPZ.

\begin{figure}
\centering
\includegraphics[width=\columnwidth]{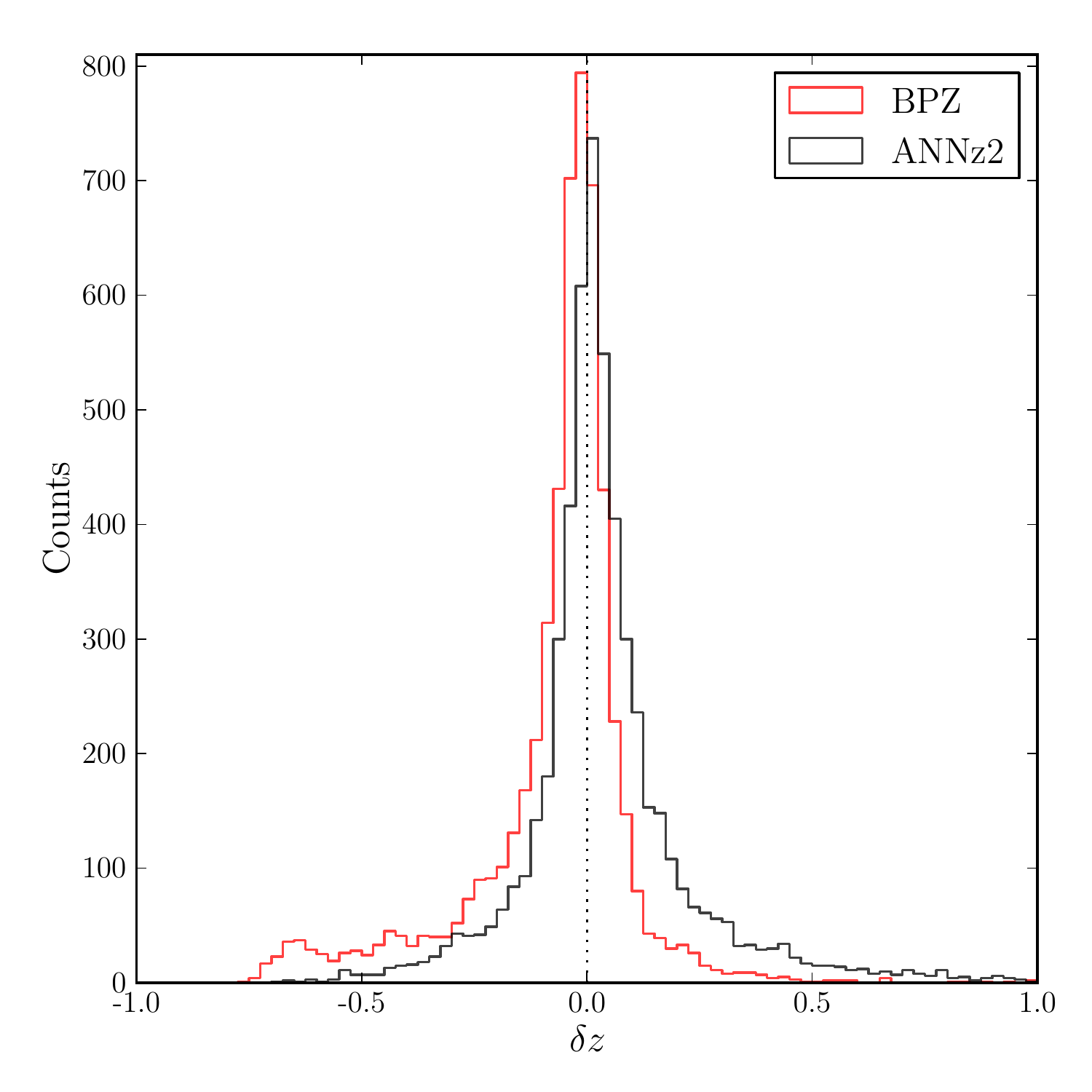}
\caption{Comparison of normalised photometric redshift errors for
  ANNz2 (black) and BPZ (red) in the CDFS field.}\label{ANNz2bias}
\end{figure}

A more relevant test in view of future applications of the KiDS
machine-learning photo-z's at the full depth of the survey is by
selecting deep spectroscopic samples for tests. For this purpose we
used the COSMOS-KiDS ($\overline{z} =0.706$) and CDFS-KiDS
($\overline{z} =0.736$) data as two independent test sets by
training ANNz2 on spectroscopic data with each of them removed in turn
(i.e.\ training on data without COSMOS and testing on COSMOS, and
similarly for CDFS). We note that these tests might be prone to cosmic
variance as the two datasets cover only one KiDS tile each and include
just $\sim17\,000$ and $\sim5\,500$ spectroscopic sources, respectively. In this
procedure we adopted the aforementioned weighting of the training
data, as implemented in the ANNz2 code. We compare the results of these
experiments to BPZ, which does not depend on a training set.
In this case the results point to comparable performance of ANNz2 and BPZ,
although the former seems to be less biased overall. For the COSMOS
data used as the independent test set, we obtained $\overline{\delta z} = -0.0009$, $\sigma = 0.164$, NMAD $= 0.077$ and 20.8\% outliers for
ANNz2 (cf. $\overline{\delta z} = -0.066$, $\sigma = 0.173$, NMAD $= 0.078$, and 21.3\% outliers for BPZ), and when CDFS was used in the
same way, we found $\overline{\delta z} = -0.043$, $\sigma = 0.194$, NMAD = $0.096$ and 25.7\% outliers for
ANNz2 (cf. $\overline{\delta z} = -0.070$, $\sigma = 0.181$, NMAD $= 0.082$, and 23.4\% outliers for BPZ). Figure
\ref{ANNz2bias} compares the relative error $\delta z$ of ANNz2 and
BPZ in the CDFS test field, indicating that the two methods exhibit
different types of systematics. 

We make available a catalogue of photometric redshifts derived with
the ANNz2 method for all the DR3 sources that have valid GAaP
magnitudes in all four bands. Table \ref{Tab:annz2-photz} in Appendix
\ref{App:mlphotz} lists the included columns. This includes almost 39.2 million
sources, which is 80\% of the full DR3 multi-band dataset. However,
for scientific applications only part of these data will be useful and
the catalogue needs to be purified of artefacts (using appropriate
flags), stellar sources, as well as sources of unreliable
photometry. In addition, the full catalogue is deeper photometrically
than the spectroscopic training set, so to avoid often unreliable
extrapolation, the data should be preferentially limited to the depth
of the spectroscopic calibration sample. Leaving the particular
selections to the user, we provide however a `fiducial' selection for
scientific use, marked with a binary flag. The applied selection
criteria, specified in detail in Appendix \ref{App:mlphotz},
remove sources affected by artifacts, classified as stars, as well as sources
fainter than the photometry of the spectroscopic sample. The latter
criteria are used in order to avoid extrapolation beyond what is
available in training, since sources fainter than any of this limits may have unreliable ANNz2 photo-z's and should be used with care.
These selections altogether yield approx. 20.5 million sources in the fiducial sample.

\subsection{Discussion}
\label{sec:photz-comparison}

\begin{figure*}
   \centering
   \includegraphics[width=\textwidth]{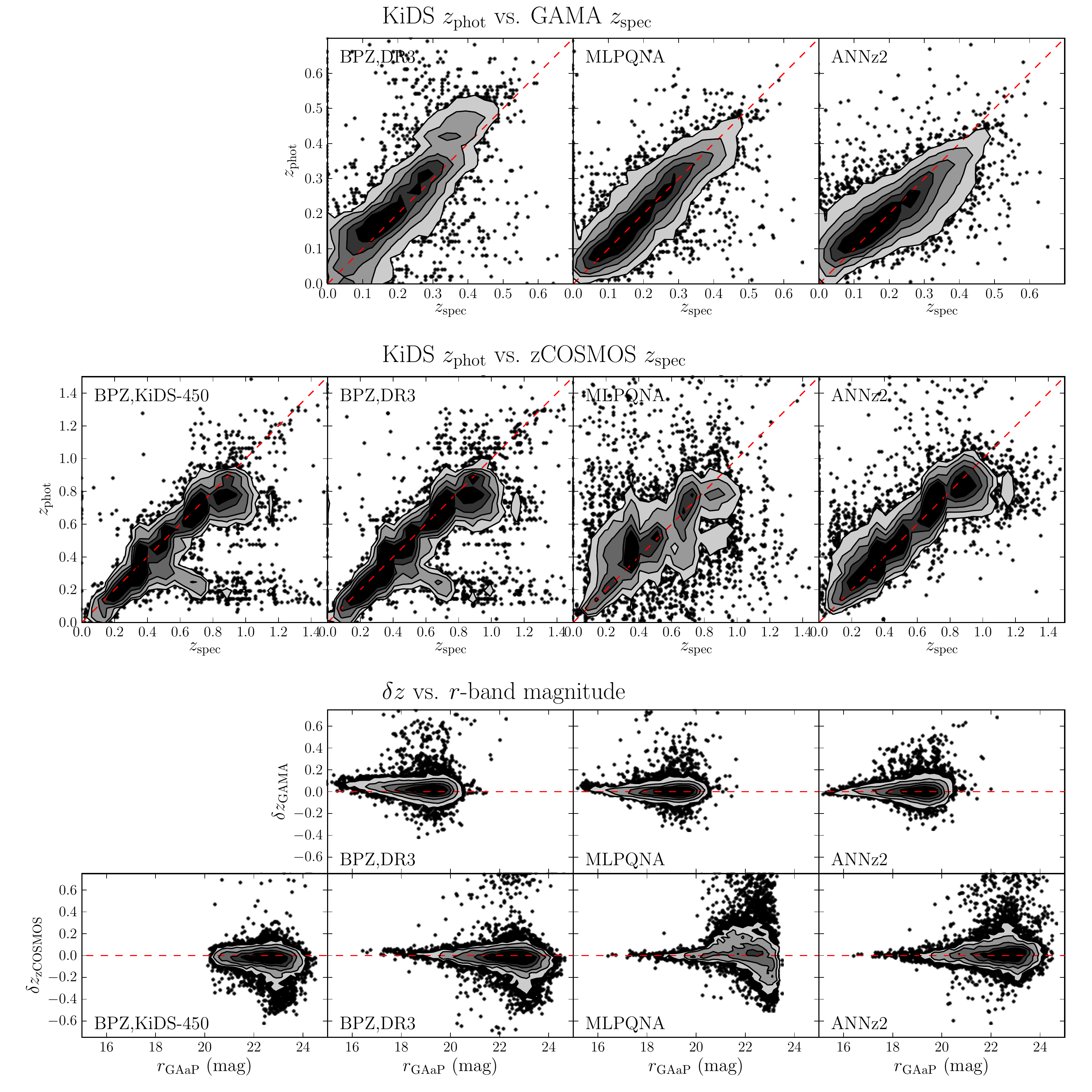}
   \caption{Comparison of photometric redshifts to spectroscopic
     redshifts. {\it Top:} direct comparisons of best-guess
     photo-z's against GAMA DR2 spec-z's. From left to right the
     panels show the BPZ photo-z's in the DR3 multi-band
     catalogue, the MLPQNA photo-z's and the ANNz2 photo-z's. The red
     dashed line corresponds to $z_\textrm{B} = z_{\textrm{spec}}$ and
     the contours are chosen randomly to enhance the clarity of the
     figures. {\it Centre:} same as the top row, but now comparing to
     zCOSMOS bright DR3 spec-z's and with on the left a panel added
     for the BPZ photo-z's included in the KiDS-450 shear
     catalogue. {\it Bottom:} normalized photo-z error $\delta z$
     versus $r$-band magnitude. The panels in the first row show the
     comparison to GAMA DR2 and the second row the comparison to
     zCOSMOS bright DR3. The order of the panels follows the order in
     the top and centre rows.}
   \label{Fig:PhotzVsSpecz}
\end{figure*}

\begin{figure*}
   \centering
   \includegraphics[width=\textwidth]{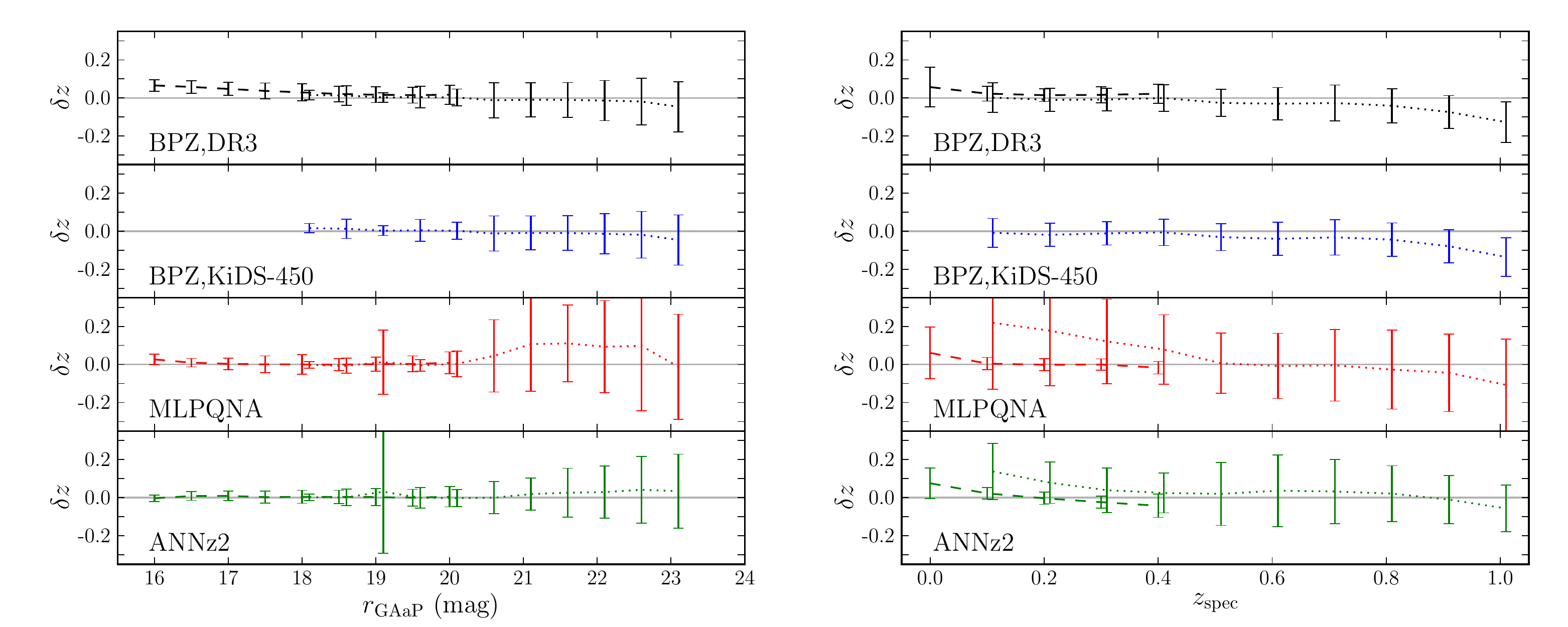}
   \caption{Trends in photometric redshift bias. {\it Left:} the
     mean normalized photo-z error $\delta z$ is plotted in bins of
     $r$-band magnitude for each of the photo-z sets. Dashed lines
     correspond to the comparison with the GAMA DR2 spec-z and the
     dotted lines to the comparison with the zCOSMOS bright DR3
     spec-z, where the latter have been offset along the x-axis
     slightly to improve clarity. Error bars show the standard
     deviation in $\delta z$. {\it Right:} same as the left panel, but
     now $\delta z$ for bins of $z_\mathrm{spec}$.} 
   \label{Fig:PhotzBiasTrends}
\end{figure*}

Due to the different methods to compute the photo-z sets described
above, including different training sets for the two machine learning
methods, an objective comparison and quality assessment is
warranted. For this purpose we juxtapose each catalogue with two publicly
available spectroscopic data sets that overlap with the KiDS-ESO-DR3
area and provide accurate redshifts in different redshift ranges.  The
second data release of GAMA \citep{liske/etal:2015} contains
spectroscopic redshifts for over 70\,000 galaxies in three 48 square
degree fields overlapping with KiDS-North. GAMA DR2 is complete to an
$r$-band Petrosian magnitude of $\sim$19, and probes redshifts up to
$\sim$0.4. The best available secure redshift estimates were obtained
from the SpecObj table, resulting in values for 70\,026 sources, of
which 76\% come from GAMA spectroscopy, 18\% from SDSS/BOSS DR10
\citep{ahn/etal:2014}, 5\% from 2dFGRS \citep{colless/etal:2001} and
the remainder from a number of other surveys.  The COSMOS field is
included in the KiDS main survey area because of the availability of
several deep photometric and spectroscopic data sets, and particularly
for the purpose of photometric redshift validation and calibration. It
should be noted that only one KiDS tile (KIDS\_150.2\_2.2) overlaps
with the COSMOS field. We use the third data release from the zCOSMOS
\citep{lilly/etal:2007} bright spectroscopic sample. From the
spectroscopic catalog we select reliable redshifts, based on both
spectroscopic and photometric information as indicated in the zCOSMOS
DR3 release notes, yielding a set of 17\,890 spec-z's.

Associating the GAMA DR2 catalogue with the DR3 BPZ, MLPQNA and ANNz2
photo-z's yields approximately 53\,000 matches, selecting only sources
classified as galaxies in the DR3 multi-band catalogue 
(\verb SG2DPHOT  = 0) 
and not masked in $r$-band 
(\verb IMAFLAGS_ISO_R  = 0). 
The bright cut-off of $r>20$ in the KiDS-450 shear catalogue
prevents association to the GAMA data. The upper row of panels in
Figure \ref{Fig:PhotzVsSpecz} shows the direct comparison of the
photo-z's and the GAMA DR2 spec-z's, while $\delta z$
(Eq.~\ref{eq:deltaz}) is plotted against $r$-band magnitude in the
third row of panels. 
Associating the KiDS photo-z catalogues to the zCOSMOS spec-z's gives
of order 10\,000 matches (Table \ref{Tab:PhotzStats}) for unmasked
galaxies\footnote{Since the KiDS-450 catalogue is already filtered to
  include only unmasked galaxies, no additional filtering is applied
  during this association}. The second row of panels in Figure
\ref{Fig:PhotzVsSpecz} shows the direct comparison between the
photo-z's and the spec-z's, and $\delta z$ is plotted against $r$-band
magnitude in the bottom row of panels\footnote{The zCOSMOS bright
  sample is magnitude limited at $I_\mathrm{AB}=22.5$, but the included
  sources extend to $\sim$24 in $r$-band GAaP magnitudes, which are
  not total magnitudes}. Statistical measures of the
photo-z biases, scatter and outlier rates are again tabulated in Table
\ref{Tab:PhotzStats} and Figure \ref{Fig:PhotzBiasTrends} shows the
trends and scatter in the photo-z bias as function of $r$-band
magnitude and spec-z.

In the magnitude/redshift regime probed by the GAMA DR2 data the
scatter and outlier rates are very similar between the three data
sets, but the machine learning methods yield significantly smaller
photo-z bias: $\overline{\delta z} = 0.02$ for BPZ versus 0.002 and
0.003 for MLPQNA and ANNz2, respectively (Table
\ref{Tab:PhotzStats}). However, the direct comparison between the ANNz2
and the GAMA DR2 data (top right panel, Fig.~\ref{Fig:PhotzVsSpecz}),
as well as the trend versus $z_\mathrm{spec}$ (bottom right panel in
Fig.~ \ref{Fig:PhotzBiasTrends}), show a trend in the bias going from
$0<z_\mathrm{spec}<0.4$.

As expected, for the much fainter and consequently noisier sources
probed by the zCOSMOS bright DR3 data, the photo-z quality
deteriorates in all aspects, showing higher bias, scatter and outlier
rates (See Table \ref{Tab:PhotzStats}). The two sets of BPZ photo-z's
show very similar behaviour.  Also in this fainter sample, there is a
clear bias present in the BPZ photo-z's, but this time negative,
$\overline{\delta z} = -0.027$ and $-0.040$. Both for the GAMA and the
zCOSMOS comparison the $\delta z$ vs. $r$ plots for BPZ show a
slightly tilted sequence. This small magnitude dependence in the bias
is confirmed by the trends visible in Figure
\ref{Fig:PhotzVsSpecz}. Furthermore, in the direct comparison of both
BPZ data sets to zCOSMOS a sequence of points can be seen along
$z_\mathrm{phot,BPZ} = 0.2$ extending to higher spec-z. This is
consistent with the bias found by \cite{hildebrandt/etal:2017} in
their lowest redshift bin ($0.1<z<0.3$).  The quality of the ANNz2
results is similar to that of BPZ with a bias of comparable amplitude,
albeit with opposite sign ($\overline{\delta z} = 0.033$), and similar
NMAD and outlier rate. However, the MLPQNA results are significantly
degraded, particularly for sources with $r$>20.5, where both the bias
and the scatter show a clear jump (Fig.~\ref{Fig:PhotzBiasTrends},
left). This is presumably caused by the knowledge base used for
training the network, which is based on SDSS and GAMA and does not
extend to these faint magnitudes. Nevertheless, even with the
inclusion of deeper spectroscopic data in the knowledge base used for
the ANNz2 data set, this comparison indicates that the BPZ results are
certainly competitive with machine learning in this domain. However,
it should be kept in mind that the fact that this particular
comparison is based on a single 1 ${\rm deg}^2$ field limits its
statistical power and renders it prone to cosmic variance.

Which set of photo-z's is preferred will depend on the scientific use
case. For relatively bright and nearby galaxies ($r<20.5$; $z<0.5$)
the MLPQNA catalogue provides the most reliable redshifts. Moving to
fainter sources the BPZ and ANNz2 results are strongly preferred, but
caution has to be observed regarding biases that can be dependent on
magnitude or redshift.

\begin{table}
\caption{KiDS photo-z quality}
\label{Tab:PhotzStats}
\centering
\footnotesize
\begin{tabular}{l | c c c c c }
\hline\hline
Set & Sources & $\overline{\delta z}$ & $\sigma$ & NMAD & Outl. \\
\hline
\multicolumn{4}{c}{KiDS vs. GAMA} \\
\hline
BPZ, DR3      & 53\,282 & 0.020  & 0.044 & 0.028 & 0.8\%  \\
MLPQNA        & 53\,008 & 0.002  & 0.042 & 0.023 & 0.6\%  \\
ANNz2         & 53\,233 & 0.003  & 0.043 & 0.030 & 0.7\%  \\
\hline
\multicolumn{4}{c}{KiDS vs. zCOSMOS} \\
\hline
BPZ, DR3      & 11\,304 & $-$0.027 & 0.124 & 0.057 & 10.0\%  \\
BPZ, KiDS-450 &  9\,150 & $-$0.040 & 0.099 & 0.059 & 10.0\%  \\
MLPQNA        &  7\,560 & 0.062   & 0.266 & 0.111 & 29.5\%  \\
ANNz2         & 10\,907 & 0.033   & 0.172 & 0.065 & 10.6\%  \\
\hline
\end{tabular}
\end{table}

\section{Data access}
\label{sec:access}

There are several ways through which KiDS-ESO-DR3 and associated data
products can be accessed. An overview is presented in this section and
up-to-date information is also available at the KiDS DR3
website.

\subsection{KiDS-ESO-DR3 main release}

The data products that constitute the main DR3 release (stacked
images, weight and flag maps, and single-band source lists for 292
survey tiles, as well as a multi-band catalog for the combined DR1, DR2
and DR3 survey area of 440 survey tiles,
see Sect. \ref{sec:dr3}) are released via the ESO Science Archive, and
also accessible via Astro-WISE and the KiDS website.

\subsubsection{ESO Science Archive}

All main release data products are disseminated through the ESO
Science Archive
Facility\footnote{\url{http://archive.eso.org/cms.html}}, which
provides several interfaces and query forms.  All image stacks, weight
maps, flag maps and single-band source lists are provided on a per
tile basis via the ``Phase 3 main query form''. This interface
supports queries on several parameters, including position, object
name, filter, observation date, etc. and allows download of the
tile-based data files. Also the multi-band catalog, which is stored in
per-tile data files, is available in this manner.
A more advanced method to query the multi-band catalog is provided by
the ``Catalogue Facility query interface'', which enables users to
perform queries on any of the catalog columns, for example
facilitating selections based on area, magnitude, photo-z or shape
information. Query results can subsequently be exported to various
(single-file) formats.

\subsubsection{Astro-WISE archive}

The data products can also be retrieved from the {\sc Astro-WISE}
system \citep{begeman/etal:2013}. This data processing and management
system is used for the production of these data products and retains
the full data lineage. For scientists interested in access to various
quality controls, further analysis tools, or reprocessing of data this
access route may be convenient.
The DBviewer web service\footnote{\url{http://dbview.astro-wise.org}}
allows querying for data products and supports file downloads, viewing
of inspection plots, and data lineage browsing. Links with DBviewer
queries to complete sets of data products are compiled on the KiDS DR3
website.

\subsubsection{KiDS DR3 website}

Apart from offering an up-to-date overview of all data access routes,
the KiDS DR3
website\footnote{\url{http://kids.strw.leidenuniv.nl/DR3}} also
provides alternative ways for data retrieval and quality control.

The synoptic table presents for each observation (tile/filter) a
combination of inspection plots relating to the image and source
extraction quality, as well as links for direct downloads of the
various data FITS files. Furthermore, direct batch downloads of all
DR3 FITS files are supported by supplying \textsc{wget} input files.

\subsection{Machine-learning photometric redshifts}

Two sets of photometric redshifts derived using machine-learning 
techniques are released. Both the MLPQNA (Sect. \ref{sec:photz-mlpqna}) 
photo-z data set, which includes PDFs, and the ANNz2 (Sect. 
\ref{sec:photz-annz2}) data set are available for download solely via the 
KiDS DR3 
website\footnote{\url{http://kids.strw.leidenuniv.nl/DR3/ml-photoz.php}}.

\subsection{Lensing shear data}

The lensing shear catalogue that was used
by \cite{hildebrandt/etal:2017} (Sect. \ref{sec:wlcatalogs}) is released both through ESO and
CADC. At the ESO Science Archive Facility the "Catalog Facility
query interface" supports queries on catalogue columns or on
position, after which the output can be exported or viewed, while the
``Phase 3 main query form'' allows the (per tile) FITS files to be
retrieved.
At CADC this weak lensing catalogue can be accessed through a web
form\footnote{\url{http://www.cadc-ccda.hia-iha.nrc-cnrc.gc.ca/en/community/kidslens/query.html}},
that allows a custom selection of columns as well as filtering for any
of the quantities provided in the catalogue.  Via the "Graphical Search Tool" the $r$-band image stacks from the lensing optimized THELI reduction, as well as the weight, sum and flag images (see Sect. \ref{sec:theli}) can be retrieved. The single-exposure frames are available on request from the authors.

Finally, batch download of the catalog FITS files is also possible
from the KiDS DR3 website.

\section{Summary and outlook}
\label{sec:summary}

The Kilo-Degree Survey \citep[KiDS, ][]{dejong/etal:2013} is an ESO Public Survey at the VLT
Survey Telescope (VST) that aims to map 1500 square degrees of
extra-galactic sky in four broad-band filters ($ugri$). Together with
its sister survey VIKING \citep{edge/etal:2013}, this will result in a 9-band
optical-infrared data set with excellent depth and image
quality. While KiDS was designed as a tomographic weak lensing survey,
the data allow for a variety of scientific analyses, ranging from
cosmology, and rare object discovery to the study of Milky Way structure.

The third data release of KiDS (Sect. \ref{sec:dr3}) adds 292 survey tiles to the previously
released 148 tiles. Released data products include astrometrically and
photometrically calibrated stacked images, weight maps, flag maps and
single-band source lists for the 292 new tiles, and a multi-band
catalogue for the combined 440 survey tiles of KiDS DR1, DR2 and
DR3 (Sect. \ref{sec:content}). This catalogue for the first time contains
Gaussian Aperture and Photometry measurements (Sect. \ref{sec:gaap}), homogenization
of the photometric calibration based on Stellar Locus Regression and
Overlap Photometry (Sect. \ref{sec:phothom}) and photometric redshifts
(Sect. \ref{sec:photz-bpz}). Cross-matching the KiDS DR3 data to Gaia DR1 \citep{gaia/etal:2016}
allows an independent verification of the photometric calibration,
indicating that the calibration is stable at the $\simeq$2\% level
(Sect. \ref{sec:gaia}).

Also made publicly available are a number of associated data
products. Accurate galaxy ellipticities are provided in a weak lensing 
shear catalogue (Sect. \ref{sec:wlcatalogs}), which was used for the
first cosmic shear measurements with KiDS data
\citep{hildebrandt/etal:2017}. The shear measurements are based on
lensing-optimized single-exposure $r$-band images (Sect. 
\ref{sec:theli}). Photometric redshifts derived using two different
Machine Learning techniques, namely Multi Layer Perceptron with Quasi 
Newton Algorithm (Sect. \ref{sec:photz-mlpqna}) and ANNz2  (Sect. 
\ref{sec:photz-annz2}), which particularly at $z<0.5$ provides more 
accurate redshifts than the template-fitting redshifts included in the 
main DR3 catalogue (see Sect. \ref{sec:photz-comparison}). 

Looking forward to the fourth data release of KiDS, we anticipate a
number of important new developments. Improved accuracy and stability of the
photometry is foreseen, as an improved calibration strategy that makes
use of the Gaia data is actively pursued. Furthermore, combining the
optical KiDS data with the near-infrared VIKING data is nearing
completion, which will allow better constrained photometric
redshifts at $z>1$. Recently the observational progress has
  improved significantly. DR4 is expected to increase the total number of released
survey tiles by approximately 250, while based on the current data rates, completion of the full
survey is anticipated during the course of 2019.

\begin{acknowledgements}
We thank L\'eon Koopmans for insightful comments and feedback.

Based on data products from observations made with ESO Telescopes at
the La Silla Paranal Observatory under programme IDs 177.A-3016,
177.A-3017 and 177.A-3018, and on data products produced by
Target/OmegaCEN, INAF-OACN, INAF-OAPD and the KiDS production team, on
behalf of the KiDS consortium. OmegaCEN and the KiDS production team
acknowledge support by NOVA and NWO-M grants. Members of INAF-OAPD and
INAF-OACN also acknowledge the support from the Department of Physics
\& Astronomy of the University of Padova, and of the Department of
Physics of Univ. Federico II (Naples). This work is supported by the 
Deutsche Forschungsgemeinschaft in the framework of the TR33 `The Dark 
Universe'. The research leading to these results has received funding 
from the People Programme (Marie Curie Actions) of the European Union's 
Seventh Framework Programme (FP7/2007-2013) under REA grant agreement 
number 627288. GAMA is a joint
European-Australasian project based around a spectroscopic campaign
using the Anglo-Australian Telescope. The GAMA input catalogue is
based on data taken from the Sloan Digital Sky Survey and the UKIRT
Infrared Deep Sky Survey. Complementary imaging of the GAMA regions is
being obtained by a number of independent survey programmes including
GALEX MIS, VST KiDS, VISTA VIKING, WISE, Herschel-ATLAS, GMRT and
ASKAP providing UV to radio coverage. GAMA is funded by the STFC (UK),
the ARC (Australia), the AAO, and the participating institutions. The
GAMA website is http://www.gama-survey.org/ .Based on zCOSMOS
observations carried out using the Very Large Telescope at the ESO
Paranal Observatory under Programme ID: LP175.A-08392. dFLenS is based
on data acquired through the Australian Astronomical Observatory,
under program A/2014B/008.

JTAdJ is supported by the Netherlands Organisation for Scientific
Research (NWO) through grant 614.061.610. GVK acknowledges financial
support from the Netherlands Research School for Astronomy (NOVA) and
Target. Target is supported by Samenwerkingsverband Noord Nederland,
European fund for regional development, Dutch Ministry of economic
affairs, Pieken in de Delta, Provinces of Groningen and Drenthe. HHi
is supported by an Emmy Noether grant (No. Hi 1495/2-1) of the
Deutsche Forschungsgemeinschaft. KK acknowledges support by the
Alexander von Humboldt Foundation. MBr acknowledges financial
contribution from the agreement ASI/INAF I/023/12/1. MBi is supported
by the Netherlands Organization for Scientific Research, NWO, through
grant number 614.001.451, and by the European Research Council through
FP7 grant number 279396. RN acknowledges support from the German
Federal Ministry for Economic Affairs and Energy (BMWi) provided via
DLR under project no. 50QE1103. CT is supported through an NWO-VICI
grant (project number 639.043.308). CB acknowledges the support of the
Australian Research Council through the award of a Future
Fellowship. IFC acknowledges the use of computational facilities
procured through the European Regional Development Fund, Project
ERDF-080 "A supercomputing laboratory for the University of Malta". RH
acknowledges support from the European Research Council FP7 grant
number 279396. CH acknowledges support from the European Research
Council under grant number 647112. HHo acknowledges support from the
European Research Council under FP7 grant number 279396. LM
acknowledges support from STFC grant ST/N000919/1. MV acknowledges
support from the European Research Council under FP7 grant number
279396 and the Netherlands Organisation for Scientific Research (NWO)
through grants 614.001.103. \end{acknowledgements}

\bibliographystyle{aa}
\bibliography{kids-dr3-final}

\onecolumn
\begin{appendix}

\section{Catalogue columns}
\subsection{Single-band source list columns}
\label{App:singleband}

Table \ref{Tab:singlebandcolumns} lists the columns that are present
in the single-band source lists provided in KiDS-ESO-DR3. In the following we provide additional information on certain columns:
\begin{itemize}
\item
\verb 2DPHOT:  \textsc{KiDS-CAT} star/galaxy classification bitmap based on the source morphology \citep[see ][]{dejong/etal:2015}. Values are: 1 = high confidence star candidate; 2 = unreliable source (e.g. cosmic ray); 4 = star according to star/galaxy separation criteria; 0 = all other sources (e.g. including galaxies). Sources identified as stars can thus have a flag value of 1, 4 or 5.
\item
\verb IMAFLAGS_ISO:  Bitmap of mask flags indicating the types of masked areas that intersect with each source's isophotes, as identified by the \textsc{Pulecenella} software \citep{dejong/etal:2015}. Different flag values indicate different types of areas: 1 = readout spike; 2 = saturation core; 4 = diffraction spike; 8 = primary reflection halo; 16 = secondary reflection halo; 32 = tertiary reflection halo; 64 = bad pixel.
\item
\verb FLUX_APER_*  and \verb FLUXERR_APER_*:  aperture flux measurements are included for 27 different aperture sizes. In the table only the smallest (2 pixels or 0.4\arcsec\ diameter) and the largest (200 pixels or 40\arcsec\ diameter) are listed; the label for the aperture of 28.5 pixels is FLUX\_APER\_28p5.
\end{itemize}

\begin{center}
\begin{longtable}{llll}
\caption{\label{Tab:singlebandcolumns} Columns provided in the single-band source lists.}\\
\hline\hline
Label & Format & Unit & Description \\
\hline
\endfirsthead

\multicolumn{4}{c}{\tablename\ \thetable{} -- continued from previous page}\\
\hline\hline
Label & Format & Unit & Description \\
\hline
\endhead

\hline
\multicolumn{4}{r}{{Continued on next page}}\\
\endfoot

\hline
\endlastfoot

2DPHOT & J & Source & classification (see section on star/galaxy separation)\\
X\_IMAGE & E & pixel & Object position along x      \\
Y\_IMAGE & E & pixel & Object position along y      \\
NUMBER & J &  & Running object number        \\
CLASS\_STAR & E &  & {\sc SExtractor} S/G classifier        \\
FLAGS & J &  & Extraction flags         \\
IMAFLAGS\_ISO & J &  & FLAG-image flags summed over the iso. profile\\
NIMAFLAG\_ISO & J &  & Number of flagged pixels entering IMAFLAGS\_ISO\\
FLUX\_RADIUS & E & pixel & Fraction-of-light radii        \\
KRON\_RADIUS & E & pixel & Kron apertures in units of A or B  \\
FWHM\_IMAGE & E & pixel & FWHM assuming a gaussian core     \\
ISOAREA\_IMAGE & J & pixel$^2$ & Isophotal area above Analysis threshold     \\
ELLIPTICITY & E &  & 1 - B\_IMAGE/A\_IMAGE        \\
THETA\_IMAGE & E & deg & Position angle (CCW/x)       \\
MAG\_AUTO & E & mag & Kron-like elliptical aperture magnitude      \\
MAGERR\_AUTO & E & mag & RMS error for AUTO magnitude     \\
ALPHA\_J2000 & D & deg & Right ascension of barycenter (J2000)     \\
DELTA\_J2000 & D & deg & Declination of barycenter (J2000)      \\
FLUX\_APER\_2 & E & count & Flux vector within circular aperture of 2 pixels  \\
... & ... & ... & ...         \\
FLUX\_APER\_200 & E & count & Flux vector within circular aperture of 200 pixels   \\
FLUXERR\_APER\_2 & E & count & RMS error vector for flux within aperture of 2 pixels  \\
... & ... & ... & ...         \\
FLUXERR\_APER\_200 & E & count & RMS error vector for flux within aperture of 200 pixels  \\
MAG\_ISO & E & mag & Isophotal magnitude        \\
MAGERR\_ISO & E & mag & RMS error for isophotal magnitude     \\
MAG\_ISOCOR & E & mag & Corrected isophotal magnitude       \\
MAGERR\_ISOCOR & E & mag & RMS error for corrected isophotal magnitude    \\
MAG\_BEST & E & mag & Best of MAG\_AUTO and MAG\_ISOCOR     \\
MAGERR\_BEST & E & mag & RMS error for MAG\_BEST      \\
BACKGROUND & E & count & Background at centroid position      \\
THRESHOLD & E & count & Detection threshold above background      \\
MU\_THRESHOLD & E & arcsec$^{-2}$ & Detection threshold above background      \\
FLUX\_MAX & E & count & Peak flux above background      \\
MU\_MAX & E & arcsec$^{-2}$ & Peak surface brightness above background     \\
ISOAREA\_WORLD & E & deg$^2$ & Isophotal area above Analysis threshold     \\
XMIN\_IMAGE & J & pixel & Minimum x-coordinate among detected pixels     \\
YMIN\_IMAGE & J & pixel & Minimum y-coordinate among detected pixels     \\
XMAX\_IMAGE & J & pixel & Maximum x-coordinate among detected pixels     \\
YMAX\_IMAGE & J & pixel & Maximum y-coordinate among detected pixels     \\
X\_WORLD & D & deg & Baryleft position along world x axis    \\
Y\_WORLD & D & deg & Baryleft position along world y axis    \\
XWIN\_IMAGE & E & pixel & Windowed position estimate along x     \\
YWIN\_IMAGE & E & pixel & Windowed position estimate along y     \\
X2\_IMAGE & D & pixel$^2$ & Variance along x       \\
Y2\_IMAGE & D & pixel$^2$ & Variance along y       \\
XY\_IMAGE & D & pixel$^2$ & Covariance between x and y     \\
X2\_WORLD & E & deg$^2$ & Variance along X-WORLD (alpha)      \\
Y2\_WORLD & E & deg$^2$ & Variance along Y-WORLD (delta)      \\
XY\_WORLD & E & deg$^2$ & Covariance between X-WORLD and Y-WORLD     \\
CXX\_IMAGE & E & pixel$^{-2}$ & Cxx object ellipse parameter      \\
CYY\_IMAGE & E & pixel$^{-2}$ & Cyy object ellipse parameter      \\
CXY\_IMAGE & E & pixel$^{-2}$ & Cxy object ellipse parameter      \\
CXX\_WORLD & E & deg$^{-2}$ & Cxx object ellipse parameter (WORLD units)    \\
CYY\_WORLD & E & deg$^{-2}$ & Cyy object ellipse parameter (WORLD units)    \\
CXY\_WORLD & E & deg$^{-2}$ & Cxy object ellipse parameter (WORLD units)    \\
A\_IMAGE & D & pixel & Profile RMS along major axis     \\
B\_IMAGE & D & pixel & Profile RMS along minor axis     \\
A\_WORLD & E & deg & Profile RMS along major axis (WORLD units)   \\
B\_WORLD & E & deg & Profile RMS along minor axis (WORLD units)   \\
THETA\_WORLD & E & deg & Position angle (CCW/world-x)       \\
THETA\_J2000 & E & deg & Position angle (east of north) (J2000)    \\
ELONGATION & E & deg & A\_IMAGE/B\_IMAGE         \\
ERRX2\_IMAGE & E & pixel$^2$ & Variance of position along x     \\
ERRY2\_IMAGE & E & pixel$^2$ & Variance of position along y     \\
ERRXY\_IMAGE & E & pixel$^2$ & Covariance of position between x and y   \\
ERRX2\_WORLD & E & deg$^2$ & Variance of position along X-WORLD (alpha)    \\
ERRY2\_WORLD & E & deg$^2$ & Variance of position along Y-WORLD (delta)    \\
ERRXY\_WORLD & E & deg$^2$ & Covariance of position X-WORLD/Y-WORLD      \\
ERRCXX\_IMAGE & E & pixel$^{-2}$ & Cxx error ellipse parameter      \\
ERRCYY\_IMAGE & E & pixel$^{-2}$ & Cyy error ellipse parameter      \\
ERRCXY\_IMAGE & E & pixel$^{-2}$ & Cxy error ellipse parameter      \\
ERRCXX\_WORLD & E & deg$^{-2}$ & Cxx error ellipse parameter (WORLD units)    \\
ERRCYY\_WORLD & E & deg$^{-2}$ & Cyy error ellipse parameter (WORLD units)    \\
ERRCXY\_WORLD & E & deg$^{-2}$ & Cxy error ellipse parameter (WORLD units)    \\
ERRA\_IMAGE & E & pixel & RMS position error along major axis    \\
ERRB\_IMAGE & E & pixel & RMS position error along minor axis    \\
ERRA\_WORLD & E & deg & World RMS position error along major axis   \\
ERRB\_WORLD & E & deg & World RMS position error along minor axis   \\
ERRTHETA\_IMAGE & E & deg & Error ellipse position angle (CCW/x)     \\
ERRTHETA\_WORLD & E & deg & Error ellipse position angle (CCW/world-x)     \\
ERRTHETA\_J2000 & E & deg & J2000 error ellipse pos. angle (east of north)  \\
FWHM\_WORLD & E & deg & FWHM assuming a gaussian core     \\
ISO0 & J & pixel$^2$ & Isophotal area at level 0     \\
ISO1 & J & pixel$^2$ & Isophotal area at level 1     \\
ISO2 & J & pixel$^2$ & Isophotal area at level 2     \\
ISO3 & J & pixel$^2$ & Isophotal area at level 3     \\
ISO4 & J & pixel$^2$ & Isophotal area at level 4     \\
ISO5 & J & pixel$^2$ & Isophotal area at level 5     \\
ISO6 & J & pixel$^2$ & Isophotal area at level 6     \\
ISO7 & J & pixel$^2$ & Isophotal area at level 7     \\
SLID & K &  & \textsc{Astro-WISE} SourceList identifier       \\
SID & K &  & \textsc{Astro-WISE} source identifier     \\
HTM & K &  & Hierarchical Triangular Mesh (level 25)      \\
FLAG & K &  & Not used         \\
\hline
\end{longtable}
\end{center}

\subsection{Multi-band catalogue}
\label{App:multiband}

Table \ref{Tab:MultiBandColumns} lists the columns that are present in the multi-band catalog provided in KiDS-ESO-DR3. In the following we provide additional information on certain columns:
\begin{itemize}
\item
\verb SG2DPHOT:  \textsc{KiDS-CAT} star/galaxy classification bitmap based on the $r$-band source morphology \citep[see ][]{dejong/etal:2015}. Values are: 1 = high confidence star candidate; 2 = unreliable source (e.g. cosmic ray); 4 = star according to star/galaxy separation criteria; 0 = all other sources (e.g. including galaxies). Sources identified as stars can thus have a flag value of 1, 4 or 5.
\item
\verb IMAFLAGS_ISO_<filter>:  Bitmap of mask flags indicating the types of masked areas that intersect with each source's isophotes, as identified by the \textsc{Pulecenella} software \citep{dejong/etal:2015}. Different flag values indicate different types of areas: 1 = readout spike; 2 = saturation core; 4 = diffraction spike; 8 = primary reflection halo; 16 = secondary reflection halo; 32 = tertiary reflection halo; 64 = bad pixel; 128 = manually masked area (tiles released in DR1 and DR2 only).
\item
\verb MAG_GAAP_<filter>:  these magnitudes are based on Gaussian Aperture and Photometry measurements and are mainly intended for colour measurements, since they only probe the central regions of the source. They are not total magnitudes, except in the case of unresolved or point sources.
\item
\verb ZPT_OFFSET_<filter>:  Zero-point offsets for each filter based on stellar locus regression ($gri$) and overlap photometry ($ugri$) that homogenize the photometry over the survey. Please note that these offsets, as well as the extinction corrections \verb EXT_SFD_<filter> {\it have not} been applied in the fluxes and magnitudes provided in the catalog. These corrections should be applied as:
\begin{equation}
\textrm{MAG\_*\_<filter>}_{corrected} =  \textrm{MAG\_*\_<filter>} + \textrm{ZPT\_OFFSET\_<filter>} - \textrm{EXT\_SFD\_<filter>}
\end{equation}
\item
\verb ODDS:  a measure of the uni-modality of the redshift Probability Distribution Function; a higher value indicates a higher reliability of the best photo-z estimate.
\item
\verb T_B:  the best-fit spectral template for each source; these values correspond to the following types, where fractional types can occur because the templates are interpolated: 1=CWW-Ell, 2=CWW-Sbc, 3=CWW-Scd, 4=CWW-Im, 5=KIN-SB3, 6=KIN-SB2 (Capak, 2004, PhD. thesis, Univ. Hawai'i)
\item
\verb TILE_FLAG:  bitmap that identifies coadds that are severely affected by image defects such as scattered light features. Masking of these features is currently not available for the imaging data released in KiDS-ESO-DR3, so data tiles flagged with this flag should be used with caution. The flag values indicate which filter is affected: 1 = $u$; 2 = $g$; 3 = $r$; 4 = $i$.
\end{itemize}

\begin{center}
\begin{longtable}{llll}
\caption{\label{Tab:MultiBandColumns} Columns provided in the
  multi-band catalogue.}\\
\hline\hline
Label & Format & Unit & Description \\
\hline
\endfirsthead

\multicolumn{4}{c}{\tablename\ \thetable{} -- continued from previous page}\\
\hline\hline
Label & Format & Unit & Description \\
\hline
\endhead

\hline
\multicolumn{4}{r}{{Continued on next page}}\\
\endfoot

\hline
\endlastfoot
ID & 23A & & Source identifier \\
RAJ2000 & D & deg & Right ascension (J2000) \\
DECJ2000 & D & deg & Declination (J2000) \\
SG2DPHOT & K &  & Source classification  \\
A & D & pixel & Linear semi major axis   \\
B & D & pixel & Linear semi minor axis   \\
CLASS\_STAR & E &  & {\sc SExtractor} star/galaxy classifier   \\
ELLIPTICITY & E &  & Ellipticity ($1-a/b$) \\
KRON\_RADIUS & E & pixel & Kron-radius used for MAG\_AUTO  \\
POSANG & E & deg & Position angle  \\
A\_GAAP & D & arcsec & Linear semi major axis of GAaP aperture \\
B\_GAAP & D & arcsec & Linear semi minor axis of GAaP aperture \\
POSANG\_GAAP & E & deg & Position angle of GAaP aperture \\
\hline
\multicolumn{4}{c}{Measurements provided for each filter}\\
\hline
FLUX\_APER\_100\_<filter> & E & count & flux in 100 pixel aperture \\
FLUX\_APER\_10\_<filter> & E & count & flux in 10 pixel aperture \\
FLUX\_APER\_14\_<filter> & E & count & flux in 14 pixel aperture \\
FLUX\_APER\_25\_<filter> & E & count & flux in 25 pixel aperture \\
FLUX\_APER\_40\_<filter> & E & count & flux in 40 pixel aperture \\
FLUX\_APER\_4\_<filter> & E & count & flux in 4 pixel aperture \\
FLUX\_APER\_6\_<filter> & E & count & flux in 6 pixel aperture \\
FLUXERR\_APER\_100\_<filter> & E & count & flux error in 100 pixel aperture \\
FLUXERR\_APER\_10\_<filter> & E & count & flux error in 10 pixel aperture \\
FLUXERR\_APER\_14\_<filter> & E & count & flux error in 14 pixel aperture \\
FLUXERR\_APER\_25\_<filter> & E & count & flux error in 25 pixel aperture \\
FLUXERR\_APER\_40\_<filter> & E & count & flux error in 40 pixel aperture \\
FLUXERR\_APER\_4\_<filter> & E & count & flux error in 4 pixel aperture \\
FLUXERR\_APER\_6\_<filter> & E & count & flux error in 6 pixel aperture \\
FLUX\_APERCOR\_100\_<filter> & E & count & corrected flux in 100 pixel aperture \\
FLUX\_APERCOR\_10\_<filter> & E & count & corrected flux in 10 pixel aperture \\
FLUX\_APERCOR\_14\_<filter> & E & count & corrected flux in 14 pixel aperture \\
FLUX\_APERCOR\_25\_<filter> & E & count & corrected flux in 25 pixel aperture \\
FLUX\_APERCOR\_40\_<filter> & E & count & corrected flux in 40 pixel aperture \\
FLUX\_APERCOR\_4\_<filter> & E & count & corrected flux in 4 pixel aperture \\
FLUX\_APERCOR\_6\_<filter> & E & count & corrected flux in 6 pixel aperture \\
FLUX\_RADIUS\_<filter> & E & pixel & {\sc SExtractor} FLUX\_RADIUS  \\
FLUXERR\_APERCOR\_100\_<filter> & E & count & corrected flux error in 100 pixel aperture \\
FLUXERR\_APERCOR\_10\_<filter> & E & count & corrected flux error in 10 pixel aperture \\
FLUXERR\_APERCOR\_14\_<filter> & E & count & corrected flux error in 14 pixel aperture \\
FLUXERR\_APERCOR\_25\_<filter> & E & count & corrected flux error in 25 pixel aperture \\
FLUXERR\_APERCOR\_40\_<filter> & E & count & corrected flux error in 40 pixel aperture \\
FLUXERR\_APERCOR\_4\_<filter> & E & count & corrected flux error in 4 pixel aperture \\
FLUXERR\_APERCOR\_6\_<filter> & E & count & corrected flux error in 6 pixel aperture \\
FLUX\_RADIUS\_<filter> & E & pixel & {\sc SExtractor} FLUX\_RADIUS \\
FWHM\_IMAGE\_<filter> & E & pixel & {\sc SExtractor} FWHM\_IMAGE \\
FLAG\_<filter> & J & & {\sc SExtractor} extraction flag \\
IMAFLAGS\_ISO\_<filter> & J & & Mask flag \\
MAGERR\_AUTO\_<filter> & E & mag & RMS error for MAG\_AUTO  \\
MAGERR\_ISO\_<filter> & E & mag & RMS error for MAG\_ISO  \\
MAG\_AUTO\_<filter> & E & mag & Kron-like elliptical aperture magnitude\tablefootmark{a} \\
MAG\_ISO\_<filter> & E & mag & Isophotal magnitude\tablefootmark{a}   \\
NIMAFLAGS\_ISO\_<filter> & J &  & Number of masked pixels entering IMAFLAGS\_ISO  \\
ISOAREA\_IMAGE\_<filter> & J & pixel$^2$ & Isophotal aperture \\
XPOS\_<filter> & E & pixel & X pixel position <filter> coadd \\
YPOS\_<filter> & E & pixel & Y pixel position <filter> coadd \\
MAG\_GAAP\_<filter> & E & mag & GAaP magnitude\tablefootmark{a} \\
MAGERR\_GAAP\_<filter> & E & mag & error in GAaP magnitude \\
ZPT\_OFFSET\_<filter> & E & mag & global photometry ZPT offset \\
EXT\_SFD\_<filter> & E & mag & Galactic foreground extinction following Schlegel et al. maps \\
\hline
\multicolumn{4}{c}{Other columns}\\
\hline
SCID & K &  & \textsc{Astro-WISE} SourceCollection identifier \\
SLID & K &  & \textsc{Astro-WISE} SourceList identifier \\
SID & K &  & \textsc{Astro-WISE} source identifier \\
Z\_B\_BPZ & E &  & Best-fitting \textsc{BPZ} photometric redshift \\
ODDS\_BPZ & E &  & Empirical ODDS of Z\_B\_BPZ \\
T\_B\_BPZ & E &  & Best-fitting BPZ spectral type\tablefootmark{b} \\
TILE\_FLAG & J &  & Tile quality warning flag \\
colour\_GAAPHOM\_U\_G & E & mag & Homogenized and Extinction corrected GAaP u-g band color \\
colour\_GAAPHOM\_U\_R & E & mag & Homogenized and Extinction corrected GAaP u-r band color \\
colour\_GAAPHOM\_U\_I & E & mag & Homogenized and Extinction corrected GAaP u-i band color \\
colour\_GAAPHOM\_G\_R & E & mag & Homogenized and Extinction corrected GAaP g-r band color \\
colour\_GAAPHOM\_G\_I & E & mag & Homogenized and Extinction corrected GAaP g-i band color \\
colour\_GAAPHOM\_R\_I & E & mag & Homogenized and Extinction corrected GAaP r-i band color \\
\hline
\end{longtable}
\tablefoot{
\tablefoottext{a}{No extinction correction or homogenization applied.}
\tablefoottext{b}{1=CWW-Ell, 2=CWW-Sbc, 3=CWW-Scd, 4=CWW-Im,
5=KIN-SB3, 6=KIN-SB2}
}
\end{center}
 
\subsection{Machine learning photometric redshifts}
\label{App:mlphotz}

\subsubsection{MLPQNA}

The best photometric redshifts derived using the MLPQNA technique (Sect. \ref{sec:photz-mlpqna}) are
provided in a single catalogue, featuring the columns listed in Table
\ref{Tab:mlpqna-photz}. Apart from the photo-z values the only content
consists of source ID information and positions that allow
straightforward association with the KiDS DR3 multi-band catalogue.

Since the knowledge base used to train the network does not span the
complete magnitude range of sources in the KiDS DR3 multi-band catalogue,
the following magnitude cuts were applied, resulting in the final set
of 8\,586\,152 photo-z's.\\
16.84 < \verb MAG_APER_20_U  < 28.55\\
16.81 < \verb MAG_APER_30_U  < 28.14\\
16.85 < \verb MAG_GAAP_U  < 28.81\\
16.18 < \verb MAG_APER_20_G  < 24.45\\
15.86 < \verb MAG_APER_30_G  < 24.59\\
16.02 < \verb MAG_GAAP_G  < 24.49\\
15.28 < \verb MAG_APER_20_R  < 23.24\\
14.98 < \verb MAG_APER_30_R  < 23.30\\
15.15 < \verb MAG_GAAP_R  < 23.29\\
14.90 < \verb MAG_APER_20_I  < 22.84\\
14.56 < \verb MAG_APER_30_I  < 23.07\\
14.75 < \verb MAG_GAAP_I  < 22.96

Also available are photo-z Probability Distribution Functions (PDFs)
based on the MLPQNA technique. These are provided in separate
catalogue files per DR3 survey tile, and their format is specified in
Table \ref{Tab:mlpqna-pdf} below. Again, included data is limited to
the PDF bins and source ID and position information that can be used
to associate to the DR3 multi-band catalogue.

\begin{center}
\begin{longtable}{llll}
\caption{\label{Tab:mlpqna-photz} Columns in the MLPQNA photo-z catalogue}\\
\hline\hline
Label & Format & Unit & Description \\
\hline
\endfirsthead

\multicolumn{4}{c}{\tablename\ \thetable{} -- continued from previous page}\\
\hline\hline
Label & Format & Unit & Description \\
\hline
\endhead

\hline
\multicolumn{4}{r}{{Continued on next page}}\\
\endfoot

\hline
\endlastfoot

ID & 25A &  & Source identifier \\
SLID & J &  & \textsc{Astro-WISE} SourceList identifier \\
SID & J &  & \textsc{Astro-WISE} source identifier \\
RA & D & deg & Right ascenscion (J2000) \\
DEC & D & deg & Declination (J2000) \\
Z\_MLPQNA & D &  & Best MLPQNA predicted photometric redshift \\
\hline
\end{longtable}
\end{center}

\begin{center}
\begin{longtable}{llll}
\caption{\label{Tab:mlpqna-pdf} Columns in the MLPQNA photo-z PDF catalogue}\\
\hline\hline
Label & Format & Unit & Description \\
\hline
\endfirsthead

\multicolumn{4}{c}{\tablename\ \thetable{} -- continued from previous page}\\
\hline\hline
Label & Format & Unit & Description \\
\hline
\endhead

\hline
\multicolumn{4}{r}{{Continued on next page}}\\
\endfoot

\hline
\endlastfoot

ID & 25A &  & Source identifier \\
SLID & J &  & \textsc{Astro-WISE} SourceList identifier \\
SID & J &  & \textsc{Astro-WISE} source identifier \\
RA & D & deg & Right ascenscion (J2000) \\
DEC & D & deg & Declination (J2000) \\
PDF001 & D &  & MLPQNA photo-z PDF bin 1 ($z$=0.01) \\
... & ... & ... & ...         \\
PDF350 & D &  & MLPQNA photo-z PDF bin 350 ($z$=3.5) \\
\hline
\end{longtable}
\end{center}

\subsubsection{ANNz2}

The best photometric redshifts derived using the ANNz2 technique
(Sect. \ref{sec:photz-annz2}) are 
provided in a single catalogue, featuring the columns listed in Table
\ref{Tab:annz2-photz}. Source ID information and positions allow
straightforward association with the KiDS DR3 multi-band catalogue.
Apart from the photo-z values, a flag column is provided that
defines the `fiducial' set of reliable photo-z's. 
This flag indicates the following selections (all are merged using an 'and' condition):
\begin{itemize}
\item 
(\verb IMAFLAGS_ISO_band  \& 01010111) == 0 for each band \citep[artefact masking, following][]{radovich/etal:2017};
\item
\verb SG2DPHOT  == 0 (star removal);
\item
\verb MAGERR_GAAP_band  >0 for each band (bad photometry removal);
\item
\verb MAG_GAAP_U  <25.4 \&\& 
\verb MAG_GAAP_G  <25.6 \&\& 
\verb MAG_GAAP_R  <24.7 \&\& 
\verb MAG_GAAP_I  <24.5
\end{itemize}

The latter criteria reflect the 99.9 percentile of photometry in each the bands of the spectroscopic sample, and are applied to avoid extrapolation beyond what is available in training. In other words, sources fainter than any of these limits may have unreliable ANNz2 photo-z's and should be used with care.
These selections altogether yield approx. 20.5 million sources in the fiducial sample.

\begin{center}
\begin{longtable}{llll}
\caption{\label{Tab:annz2-photz} Columns in the ANNz2 photo-z catalogue}\\
\hline\hline
Label & Format & Unit & Description \\
\hline
\endfirsthead

\multicolumn{4}{c}{\tablename\ \thetable{} -- continued from previous page}\\
\hline\hline
Label & Format & Unit & Description \\
\hline
\endhead

\hline
\multicolumn{4}{r}{{Continued on next page}}\\
\endfoot

\hline
\endlastfoot

ID & 23A &  & Source identifier \\
RAJ2000 & D & deg & Right ascenscion (J2000) \\
DECJ2000 & D & deg & Declination (J2000) \\
zphot\_ANNz2 & D &  & Best ANNz2 predicted photometric redshift \\
fiducial & I &  & Flag defining fiducial selection \\
\hline
\end{longtable}
\end{center}

\subsection{KiDS-450 shear catalogue}
\label{App:shearcat}

Table \ref{Tab:ShearCatColumns} lists the columns provided in the KiDS-450 or KiDS-DR3.1 lensing shear catalogue. Masking of bright stars, satellite trails and other image defects results in an effective area of 360.3 ${\rm deg}^2$. The released catalogue has already been cleaned from unreliable sources, following the rejection strategy detailed in \cite{hildebrandt/etal:2017}. The faint limit of the catalogue is $r$=25.0, at an approximate signal-to-noise of 5$\sigma$. Please note that the source ID's in this catalogue are not always identical to those in the DR3 multi-band catalogue, due to the independent astrometric solution used for the lensing analysis.

\cite{hildebrandt/etal:2017} found in their cosmic shear analysis that small residual c-terms (non-zero average shear) are present in the catalogue. These were subtracted per patch and per tomographic bin before computing the shear-shear correlation functions. Any science analysis for which c-terms are important should determine these from the data, using the \textit{lens}fit weights to compute the averages.

For a number of columns additional information is provided:
\begin{itemize}
\item
\verb KIDS_TILE:  name of the KiDS survey tile in which the source is located. Searching the ESO archive with this OBJECT name will link to further data products for this tile.
\item
\verb THELI_NAME:  name for the survey tile in the \textsc{THELI} pipeline; for scripting reasons the `.' and `-' characters are replaced with `p' and `m', respectively.
\item
\verb MASK:  bit mask indicating sources affected by different types of defects. Automatic and manual masks produced during data processing flag areas affected by bright stars (e.g. saturated pixels, reflection halos, diffraction and readout spikes) and other severe image defects. In the released catalogue the most strongly affected regions are already removed, leaving only sources with reliable measurements. As a result, only the bit mask values 2 (faint stellar reflection halo), 64 ($u$-band \textsc{Astro-WISE} manual mask), 128 ($g$-band \textsc{Astro-WISE} manual mask), 256 ($r$-band \textsc{Astro-WISE} manual mask) and 512 ($i$-band \textsc{Astro-WISE} manual mask) are present. 
\item
\verb Flag:  \textsc{SExtractor} extraction flag. Many sources that are flagged during source detection are removed from the catalogue based on the \textit{lens}fit results because they do not provide reliable shear measurements. The flag values that are still present in the catalogue are 1 (the object has neighbors, bright and close enough to significantly bias the photometry, or bad pixels (more than 10\% of the integrated area affected), 2 (the object was originally blended with another one) and 16 (objects aperture data are incomplete or corrupted).
\item
\verb SG_FLAG:  star-galaxy separator based on analysis of the second and fourth order image moments of the source; 0 = star, 1 = galaxy
\item
\verb MAG_u/g/r/i:  the magnitudes are based on Gaussian Aperture and Photometry measurements and are dereddened and colour-calibrated using stellar locus regression. Note: these aperture magnitudes are mainly intended for colour measurements, since they only probe the central regions of the source. They are not total magnitudes, except in the case of unresolved or point sources.
\item
\verb MAG_LIM_u/g/r/i:  local limiting magnitude, defined as the magnitude corresponding to a flux equal to the 1$\sigma$ flux error.
\item
\verb ZPT_offset:  the magnitudes reported in this catalogue have been colour-calibrated, but their absolute calibration has only been homogenized per survey tile, not over the full area. Based on a comparison of the $r$-band magnitudes with the Gaia DR1 \citep{gaia/etal:2016} we provide these additional photometric offsets that can be used to homogenize the photometry over the whole catalogue. The reported offsets are with respect to the Gaia photometry, but can be used to calibrate the photometry to the SDSS photometric system as follows:
\begin{equation}
mag\_u/g/r/i\_homogenized = \verb mag_u/g/r/i - \verb ZPT_offset + 0.049
\end{equation}
Note: if used, these offsets must be applied to the magnitudes in all filters!
\item
\verb T_B:  the best-fit spectral template for each source; these values correspond to the following types, where fractional types can occur because the templates are interpolated: 1=CWW-Ell, 2=CWW-Sbc, 3=CWW-Scd, 4=CWW-Im, 5=KIN-SB3, 6=KIN-SB2 (Capak, 2004, PhD. thesis, Univ. Hawai'i).
\item
\verb ODDS:  a measure of the uni-modality of the redshift Probability Distribution Function; a higher value indicates a higher reliability of the best photo-z estimate.
\item
\verb fitclass:  \textit{lens}fit object class; the only classes included in the catalogue are 0 (galaxy, no issues) and -9 (large galaxy, overfills 48 pixel postage stamp size). The latter class is retained to avoid ellipticity selection bias in the brightest galaxy sample.
\item
\verb n_exposures_used:  the number of r-band sub-exposures for which \textit{lens}fit measured the shape of the source. Due to the dither pattern, or near tile edges, some sources are only present in a subset of the 5 sub-exposures.
\item
\verb e1  or \verb e2:   \textit{lens}fit shear estimators. Note: the \verb e2  component is defined relative to the RAJ2000, DECJ2000 grid; depending on the user's definition of angles in this reference frame, the sign of \verb e2 may need to be changed.
\item
\verb PSF_e1  and \verb PSF_e1_exp[k]:  model PSF ellipticities at the location of the object, in this case the real part of $\epsilon_{PSF}$. \verb PSF_e1_exp[k]  is the PSF ellipticity in sub-exposure number k, while \verb PSF_e1  refers to the average over all exposures used.
\item
\verb m: the multiplicative shear calibration correction which should be applied in an ensemble average, rather than on a galaxy-by-galaxy basis \citep[see ][]{fenechconti/etal:2016}. Averaged catalogued ellipticities should be divided by 1+<\verb m >, where the ellipticities should be weighted with the \textit{lens}fit weight.
\end{itemize}

\begin{center}
\begin{longtable}{llll}
\caption{\label{Tab:ShearCatColumns} Columns provided in the weak lensing shear catalogue.}\\
\hline\hline
Label & Format & Unit & Description \\
\hline
\endfirsthead

\multicolumn{4}{c}{\tablename\ \thetable{} -- continued from previous page}\\
\hline\hline
Label & Format & Unit & Description \\
\hline
\endhead

\hline
\multicolumn{4}{r}{{Continued on next page}}\\
\endfoot

\hline
\endlastfoot
ID                      & 25A    &        & Source identifier                                                               \\
RAJ2000                 & 1D     & deg    & Right ascension of barycenter (J2000)                                           \\
DECJ2000                & 1D     & deg    & Declination of barycenter (J2000)                                               \\
Patch                   & 3A     &        & Patch (G9, G12, G15, G23 or GS)                                                 \\
SeqNr                   & 1J     &        & Running object number within the patch                                          \\
KIDS\_TILE               & 16A    &        & Name of survey tile                                                             \\
THELI\_NAME              & 16A    &        & \textsc{THELI} name for the tile                                                         \\
MASK                    & 1J     &        & Mask value at the object position                                               \\
SG\_FLAG                 & 1E     &        & Star-galaxy separator (0=star, 1=galaxy)                                        \\
KRON\_RADIUS             & 1E     & pixel  & Scaling radius of the ellipse for magnitude measurements                     \\
Xpos                    & 1E     & pixel  & Object position along x in the $r$-band \textsc{THELI} stack (non unique)         \\
Ypos                    & 1E     & pixel  & Object position along y in the $r$-band \textsc{THELI} stack (non unique)         \\
FWHM\_IMAGE              & 1E     & pixel  & $r$-band FWHM assuming a Gaussian core                                            \\
FWHM\_WORLD              & 1E     & deg    & $r$-band FWHM assuming a Gaussian core                                            \\
Flag                    & 1J     &        & $r$-band \textsc{SExtractor} extraction flags                                              \\
FLUX\_RADIUS             & 1E     & pixel  & $r$-band half-light radius                                                        \\
CLASS\_STAR              & 1E     &        & $r$-band \textsc{SExtractor} S/G classifier output                                         \\
MAG\_u                   & 1E     & mag    & Magnitude in the $u$-band (GAaP, dereddened)\tablefootmark{a}     \\
MAGERR\_u                & 1E     & mag    & Magnitude error in the $u$-band                                                   \\
MAG\_g                   & 1E     & mag    & Magnitude in the $g$-band (GAaP, dereddened)\tablefootmark{a}  \\
MAGERR\_g                & 1E     & mag    & Magnitude error in the $g$-band                                                   \\
MAG\_r                   & 1E     & mag    & Magnitude in the $r$-band (GAaP, dereddened)\tablefootmark{a}    \\
MAGERR\_r                & 1E     & mag    & Magnitude error in the $r$-band                                                   \\
MAG\_i                   & 1E     & mag    & Magnitude in the $i$-band (GAaP, dereddened)\tablefootmark{a}      \\
MAGERR\_i                & 1E     & mag    & Magnitude error in the $i$-band                                                   \\
MAG\_LIM\_u               & 1E     & mag    & Limiting magnitude in the $u$-band                                                \\
MAG\_LIM\_g               & 1E     & mag    & Limiting magnitude in the $g$-band                                                \\
MAG\_LIM\_r               & 1E     & mag    & Limiting magnitude in the $r$-band                                                \\
MAG\_LIM\_i               & 1E     & mag    & Limiting magnitude in the $i$-band                                                \\
ZPT\_offset              & 1E     & mag    & Zeropoint offset derived from Gaia DR1 $G$ photometry                             \\
Z\_B                     & 1E     &        & \textsc{BPZ} best redshift estimate                                                      \\
Z\_B\_MIN                 & 1E     &        & Lower bound of the 95\% confidence interval of Z\_B                               \\
Z\_B\_MAX                 & 1E     &        & Upper bound of the 95\% confidence interval of Z\_B                               \\
T\_B                     & 1E     &        & Spectral type corresponding to Z\_B\tablefootmark{b}                                           \\
ODDS                    & 1E     &        & Empirical ODDS of Z\_B                                                           \\
fitclass                & 1I     &        & \textit{lens}fit:  fit class                                                              \\
bias\_corrected\_scalelength & 1E     & pixel     & \textit{lens}fit:  galaxy model scale length                                     \\
bulge\_fraction          & 1E     &        & \textit{lens}fit:  galaxy model bulge-fraction B/T                                        \\
model\_flux              & 1E     & counts & \textit{lens}fit:  galaxy model flux                                                      \\
pixel\_SNratio           & 1E     &        & \textit{lens}fit:  data S/N ratio                                                         \\
model\_SNratio           & 1E     &        & \textit{lens}fit:  model S/N ratio                                                        \\
contamination\_radius    & 1E     & pixel  & \textit{lens}fit:  distance to nearest contaminating isophote                     \\
PSF\_e1                  & 1E     &        & \textit{lens}fit:  PSF model mean ellipticity e1                                          \\
PSF\_e2                  & 1E     &        & \textit{lens}fit:  PSF model mean ellipticity e2                                          \\
PSF\_Strehl\_ratio        & 1E     &        & \textit{lens}fit:  PSF model mean pseudo-Strehl ratio                                 \\
PSF\_Q11                 & 1E     &        & \textit{lens}fit:  2nd order brightness moment Q11 of the PSF                        \\
PSF\_Q22                 & 1E     &        & \textit{lens}fit:  2nd order brightness moment Q22 of the PSF                        \\
PSF\_Q12                 & 1E     &        & \textit{lens}fit:  2nd order brightness moment Q12 of the PSF                        \\
n\_exposures\_used        & 1E     &        & \textit{lens}fit:  number of r-band exposures used in \textit{lens}fit measurements \\
PSF\_e1\_exp1             & 1E     &        & \textit{lens}fit:  PSF model ellipticity e1 of exposure 1                                 \\
PSF\_e2\_exp1             & 1E     &        & \textit{lens}fit:  PSF model ellipticity e2 of exposure 1                                 \\
PSF\_e1\_exp2             & 1E     &        & \textit{lens}fit:  PSF model ellipticity e1 of exposure 2                                 \\
PSF\_e2\_exp2             & 1E     &        & \textit{lens}fit:  PSF model ellipticity e2 of exposure 2                                 \\
PSF\_e1\_exp3             & 1E     &        & \textit{lens}fit:  PSF model ellipticity e1 of exposure 3                                 \\
PSF\_e2\_exp3             & 1E     &        & \textit{lens}fit:  PSF model ellipticity e2 of exposure 3                                 \\
PSF\_e1\_exp4             & 1E     &        & \textit{lens}fit:  PSF model ellipticity e1 of exposure 4                                 \\
PSF\_e2\_exp4             & 1E     &        & \textit{lens}fit:  PSF model ellipticity e2 of exposure 4                                 \\
PSF\_e1\_exp5             & 1E     &        & \textit{lens}fit:  PSF model ellipticity e1 of exposure 5                                 \\
PSF\_e2\_exp5             & 1E     &        & \textit{lens}fit:  PSF model ellipticity e2 of exposure 5                                 \\
e1                      & 1E     &        & \textit{lens}fit:  galaxy e1 expectation value                                            \\
e2                      & 1E     &        & \textit{lens}fit:  galaxy e2 expectation value                                            \\
weight                  & 1E     &        & \textit{lens}fit:  inverse variance shear weight                                          \\
m                       & 1E     &        & Multiplicative shear calibration                                                \\
\hline
\end{longtable}
\tablefoot{
\tablefoottext{a}{Absolute calibration via ZPT\_offset not applied.}
\tablefoottext{b}{1=CWW-Ell, 2=CWW-Sbc, 3=CWW-Scd, 4=CWW-Im,
5=KIN-SB3, 6=KIN-SB2}
}
\end{center}

\end{appendix}

\end{document}